\documentclass[10pt,conference,final]{IEEEtran}
\IEEEoverridecommandlockouts

\usepackage{cite}
\usepackage{amsmath,amsthm,amssymb,stmaryrd}
\usepackage{url}
\usepackage{cleveref}
\usepackage[pdftex,all]{xy}
\SelectTips{cm}{}
\usepackage{xcolor}
\usepackage[pdftex]{graphicx}
\usepackage{mathtools}
\usepackage{tikz}
\usetikzlibrary{matrix}
\usetikzlibrary{positioning}
\usepackage{ifthen}
\usepackage{comment}

\specialcomment{auxproof}
{\mbox{}\newline\textbf{BEGIN: AUX-PROOF}\dotfill\newline}
{\mbox{}\newline\textbf{END: AUX-PROOF}\dotfill\newline}

\theoremstyle{definition}
\newtheorem{mydefinition}{Definition}[section]
\theoremstyle{plain}
\newtheorem{mylemma}[mydefinition]{Lemma}
\newtheorem{myproposition}[mydefinition]{Proposition}
\newtheorem{mytheorem}[mydefinition]{Theorem}

\newtheorem{mycorollary}[mydefinition]{Corollary}

\theoremstyle{definition}

\newtheorem{myexample}[mydefinition]{Example}

\newtheorem{mynotation}[mydefinition]{Notation}

\theoremstyle{remark}
\newtheorem*{myproof}{Proof}
\def\myqed{\qed}

\newcommand{\place}{\underline{\phantom{n}}\,} 

\newcommand{\op}{\mathrm{op}}

\newcommand{\id}{\mathrm{id}}

\newcommand{\pow}{\mathcal{P}}

\newcommand{\sdist}{\mathcal{D}_{\le 1}}

\newcommand{\CLatw}{\mathbf{CLat}_{\sqcap}}

\newcommand{\Set}{\mathbf{Set}}

\newcommand{\Pre}{\mathbf{Pre}}

\newcommand{\Top}{\mathbf{Top}}
\newcommand{\PMet}{\mathbf{PMet}_{1}}

\newcommand{\Meas}{\mathbf{Meas}}

\newcommand{\EM}{\mathcal{E}{\kern-.5ex}\mathcal{M}}
\newcommand{\Kl}{\mathcal{K}{\kern-.2ex}\ell}

\newcommand{\AAA}{\mathbb A}
\newcommand{\BB}{\mathbb B}
\newcommand{\CC}{\mathbb C}
\newcommand{\DD}{\mathbb D}
\newcommand{\EE}{\mathbb E}
\newcommand{\FF}{\mathbb F}

\newcommand{\arrow}{\mathrel{\rightarrow}}

\newcommand{\darrow}{\mathrel{\dot\arrow}}
\newcommand{\ndarrow}{\mathrel{\dot\nrightarrow}}
\newcommand{\bOmega}{\mathbf{\Omega}}

\newcommand{\co}{\mathbin{\circ}}

\newcommand{\adjunction}[3]{
  \ar@<.4pc>[#1]^-{#2}
  \ar@{}[#1]|-*=0[@]{\bot}
  \ar@<-.4pc>@{<-}[#1]_-{#3}
}

\newcommand{\expect}[2]{E_{#1}[#2]}

\newcommand{\ERel}{\mathbf{ERel}}
\newcommand{\EqRel}{\mathbf{EqRel}}

\newcommand{\fa}[1]{\forall #1.~}
\newcommand{\ex}[1]{\exists #1.~}
\newcommand{\Eq}{\mathrm{Eq}}

\newcommand{\Duplicator}{\text{D}}
\newcommand{\Spoiler}{\text{S}}

\newcommand{\fib}[3]{#1 \xrightarrow{#2} #3}
\newcommand{\fibp}{\fib{\EE}{p}{\CC}}

\newcommand{\Lift}{{\bf Lift}}

\makeatletter
\newcommand\newtag[2]{#1\def\@currentlabel{#1}\label{#2}}
\makeatother

\renewcommand{\fib}[3]{
  \ifthenelse{\equal{#2}{}}{#1\rightarrow #3}{#2:#1\rightarrow #3}
}

\begin{document}

\title{
  Codensity Games for Bisimilarity
  \thanks{
    The authors are supported by ERATO HASUO Metamathematics
    for Systems Design Project (No.~JPMJER1603), JST.  S.K.\ and I.H.\
    are supported by the JSPS-Inria Bilateral Joint Research Project
    CRECOGI; I.H.\ is supported by Grants-in-Aid No.~15KT0012 and
    15K11984, JSPS; B.K.\ is supported by the ERC under the European
    Union's Horizon 2020 research and innovation programme (ERC
    consolidator grant LIPA, agreement no.\ 683080).  Part of the work
    was done during N.H.'s internship, and B.K.'s visit, at National
    Institute of Informatics, Tokyo, Japan.
  }
}

\author{
  \IEEEauthorblockN{
    Yuichi Komorida\IEEEauthorrefmark{1}\IEEEauthorrefmark{2},
    Shin-ya Katsumata\IEEEauthorrefmark{1},
    Nick Hu\IEEEauthorrefmark{3},
    Bartek Klin\IEEEauthorrefmark{4},
    Ichiro Hasuo\IEEEauthorrefmark{1}\IEEEauthorrefmark{2}a}
  \IEEEauthorblockA{\IEEEauthorrefmark{1} National Institute of Informatics, Tokyo, Japan}
  \IEEEauthorblockA{\IEEEauthorrefmark{2} The Graduate University for Advanced Studies (SOKENDAI), Hayama, Japan}
  \IEEEauthorblockA{\IEEEauthorrefmark{3} University of Oxford, United Kingdom}
  \IEEEauthorblockA{\IEEEauthorrefmark{4} University of Warsaw, Poland}
}

\IEEEoverridecommandlockouts
\IEEEpubid{\makebox[\columnwidth]{978-1-7281-3608-0/19/\$31.00~\copyright2019 IEEE \hfill}\hspace{\columnsep}\makebox[\columnwidth]{ }}

\maketitle

\begin{abstract}
  Bisimilarity as an equivalence notion of systems has been central
  to process theory. Due to the recent rise of interest in
  quantitative systems (probabilistic, weighted, hybrid, etc.),
  bisimilarity has been extended in various ways: notably,
  bisimulation metric between probabilistic systems. An important
  feature of bisimilarity is its game-theoretic characterization,
  where Spoiler and Duplicator play against each other; extension of
  bisimilarity games to quantitative settings has been actively
  pursued too.

  In this paper, we present a general framework that uniformly
  describes game characterizations of bisimilarity-like notions. Our
  framework is formalized categorically using fibrations and
  coalgebras. In particular, our characterization of bisimilarity in
  terms of fibrational predicate transformers allows us to derive
  codensity bisimilarity games: a general categorical game
  characterization of bisimilarity. Our framework covers known
  bisimilarity-like notions (such as bisimulation metric) as well as
  new ones (including what we call bisimulation topology).
\end{abstract}
\section{Introduction}
\subsection{Bisimilarity Notions and Games}
\label{subsec:introBisimBisimGame}

Since the seminal works by Park and Milner~\cite{Park81,Milner89},
\emph{bisimilarity} has played a central role in theoretical computer
science. It is an equivalence notion between branching systems; it
abstracts away internal states and stresses the black-box
observation-oriented view on process semantics. Bisimilarity is
usually defined as the largest \emph{bisimulation}, which is a binary relation
that satisfies a suitable mimicking condition. In fact, a
bisimulation $R$ can be characterized as a post-fixed point
$R\subseteq \Phi(R)$ using a suitable relation transformer $\Phi$;
from this we obtain that bisimilarity is the greatest \emph{fixed}
point of $\Phi$ by the Knaster--Tarski theorem. This order-theoretic
foundation is the basis of a variety of advanced techniques for
reasoning about (or using) bisimilarity, such as bisimulation
up-to---see, e.g.,~\cite{SangiorgiR11}.

Bisimilarity is conventionally defined for state-based systems with
nondeterministic branching. However, as the applications of computer
systems become increasingly pervasive and diverse (such as
cyber-physical systems), extension of bisimilarity to systems with
other branching types has been energetically sought in the
literature. One notable example is the bisimulation notion for
probabilistic systems in~\cite{LarsenS91}: it is a relation that
witnesses that two states are indistinguishable in their behaviors
henceforth. This qualitative notion has also been made quantitative,
as the notion of \emph{bisimulation
  metric}~\cite{DesharnaisGJP04}. It replaces a relation with a
metric that is induced by the probabilistic transition structure.

There is a body of literature
(including~\cite{HermidaJ98,HasuoKC18,BaldanBKK18,BonchiPPR14,KoenigM18,BonchiKP18,WissmannDKH19FoSSaCSToAppear})
that aims to identify the mathematical essences that are shared by
this variety of bisimilarity, and express the identified
essences in a rigorous manner using \emph{category theory}. Our
particular interest is in the correspondence between bisimilarity
notions and \emph{(safety) games}; three examples of which are given
below. This interest in bisimilarity games is shared by the recent
work~\cite{KoenigM18}, and the comparison is discussed
in~\S{}\ref{subsec:relatedWork}.

\subsubsection{Bisimilarity Games}
It is well-known that the following game characterizes the
conventional notion of bisimilarity between Kripke frames. Let
$(X,{\to})$ be a Kripke frame where ${\to}\subseteq X^{2}$; the game
is played between Duplicator (D) and Spoiler (S). In a position
$(x_{1}, x_{2})$, Spoiler challenges Duplicator's claim that $x_{1}$
and $x_{2}$ are bisimilar, by choosing one of the states (say
$x_{1}$) and further choosing a transition $x_{1}\to
x_{1}'$. Duplicator responds by choosing a transition
$x_{2}\to x_{2}'$ from the other state, and the game is continued
from $(x_{1}',x_{2}')$. Duplicator wins
if Spoiler gets stuck, or the game continues infinitely long, and this
witnesses that $x_{1}$ and $x_{2}$ are bisimilar.

\subsubsection{Games for Probabilistic Bisimilarity}
A recent step forward in the topic of bisimilarity and games is the
characterization of probabilistic bisimulation introduced
in~\cite{FijalkowKP17}. For simplicity, here we describe its
discrete version.

Let $(X,c)$ be a Markov chain, where $X$ is a countable set of
states, and $c\colon X\to \sdist X$ is a transition kernel that
assigns to each state $x\in X$ a probability subdistribution
$c(x)\in \sdist X$. Here
$\sdist X=\{d\colon X\to [0,1]\mid \sum_{x\in X}d(x)\le 1\}$ denotes
the set of probability subdistributions over $X$.  For
$Z\subseteq X$, let $c(x)(Z)$ denote the probability with which a
successor of $x$ is chosen from $Z$; that is,
$c(x)(Z)=\sum_{x'\in Z}c(x)(x')$. Since $c(x)$ is only a
\emph{sub}-distribution over $X$, the probability $c(x)(X)$ is
$\le 1$ rather than $=1$. The remaining probability $1-c(x)(X)$ can
be thought of as the probability of $x$ getting stuck.

Recall from~\cite{LarsenS91} that an equivalence relation
$R\subseteq X^{2}$ is a \emph{(probabilistic) bisimulation} if, for
any $(x,y)\in R$ and each $R$-closed subset $Z\subseteq X$,
$c(x)(Z)=c(y)(Z)$ holds.

\begin{table}[htbp]
  \caption{The Game for Probabilistic Bisimilarity
    from~\cite{FijalkowKP17}}
  \label{table:probabilisticBisimGameIntroFKP}
  \centering
  \begin{tabular}{l|l|l}
    position & player &  possible moves    \\\hline\hline
    $(x,y)\in X^{2}$ & S & $Z\subseteq X \text{ s.t.\ } c(x)(Z)\neq c(y)(Z)
                           $
    \\\hline
    $Z\subseteq X$ & D &
                         $                         (x',y')\in X^{2}
                         \text{ s.t.\ }
                         x'\in Z \land y'\not\in Z
                         $
  \end{tabular}
\end{table}
The game introduced in~\cite{FijalkowKP17} is in
Table~\ref{table:probabilisticBisimGameIntroFKP}.  It is shown
in~\cite{FijalkowKP17} that Duplicator is winning in the game at
$(x,y)$ if and only if $x$ and $y$ are bisimilar, in the sense
of~\cite{LarsenS91} (recalled above).  It is not hard to find an
intuitive correspondence between the game in
Table~\ref{table:probabilisticBisimGameIntroFKP} and the definition
of bisimulation~\cite{LarsenS91}: Spoiler challenges the
bisimilarity claim between $x,y$ by exhibiting $Z$ such that
$c(x)(Z)=c(y)(Z)$ is violated; Duplicator makes a counterargument by
claiming that $Z$ is in fact not bisimilarity-closed, exhibiting a pair of states
$(x',y')$ that Duplicator claims are bisimilar.

\subsubsection{Games for Probabilistic Bisimulation Metric}\label{subsubsec:introGameProbBisimMet}
Our following observation marked the beginning of the current work:
the game for (qualitative) bisimilarity for probabilistic systems
(from~\cite{FijalkowKP17},
Table~\ref{table:probabilisticBisimGameIntroFKP}) can be almost
literally adapted to (quantitative) \emph{bisimulation metric} for
probabilistic systems. This metric was first introduced
in~\cite{DesharnaisGJP04}.

For simplicity we focus on the discrete setting; we also restrict to
pseudometrics bounded by $1$.  Let $(X,c)$ be a Markov chain with a
countable state space $X$. The \emph{bisimulation metric}
$d_{(X,c)}\colon X^{2}\to [0,1]$ is defined to be the smallest
pseudometric (with respect to the pointwise order) that makes the
transition kernel
\begin{displaymath}
  c\colon (X,\,d_{(X,c)})\longrightarrow
  \bigl(\,\sdist X,\, \mathcal{K}(d_{(X,c)})\,\bigr)
\end{displaymath}
non-expansive with respect to the specified pseudometrics. Here
$\mathcal{K}(d_{(X,c)})$ is the so-called \emph{Kantorovich
  metric} over $\sdist X$  induced by the pseudometric $d_{(X,c)}$ over $X$.
It is defined as follows. For $\mu, \nu\in \sdist X$,
\begin{equation}\label{eq:KantorovichLiftingIntro}
  \mathcal{K}(d_{(X,c)})(\mu,\nu)
  =
  \sup_f
  \left|
    \expect\mu f
    -
    \expect\nu f
  \right|,
\end{equation}
where in the above sup, $f$ ranges over all non-expansive functions from
$(X,d_{(X,c)})$ to $\bigl([0,1], d_{[0,1]}\bigr)$,
$d_{[0,1]}$ denotes the usual Euclidean
metric, and
$E_\mu[f]$ is the expectation $\sum_{x\in X}f(x)\cdot\mu(x)$ of $f$
with respect to $\mu$.

Our observation is that the bisimulation metric $d_{(X,c)}$ is
characterized by the game in
Table~\ref{table:probabilisticBisimMetricGameIntro}: Duplicator is
winning at $(x,y,\varepsilon)$ if and only if
$d_{(X,c)}(x,y)\le \varepsilon$.

The game seems to be new, although its intuition is similar to the
one for Table~\ref{table:probabilisticBisimGameIntroFKP}. Note that
the formula~(\ref{eq:KantorovichLiftingIntro}) appears in the condition
of Spoiler's moves.  Spoiler challenges by exhibiting a
``predicate'' $f$ that suggests violation of the non-expansiveness of
$c$; and Duplicator makes a counterargument that $f$ is
in fact not non-expansive and thus invalid.
\begin{table}[htbp]
  \caption{The Game for (Probabilistic) Bisimulation Metric,
    Adapting~\cite{FijalkowKP17}}
  \label{table:probabilisticBisimMetricGameIntro}
  \centering
  \begin{tabular}{l|l|l}
    position & P &  possible moves    \\\hline\hline
    $(x,y,\varepsilon)$
             & S &
                   $f\colon X\to [0,1]$ \\
    $\in X^{2}\times[0,1]$ & & such that $\left|
                               \expect{c(x)}{f}                                -
                               \expect{c(y)}{f}                                \right| >\varepsilon$
    \\\hline
    $f\colon X\to [0,1]$
             & D &
                   $(x',y',\varepsilon')\in X^{2}\times[0,1]$ \\
             & & such that $\bigl|\,f(x')-f(y')\,\bigr|>\varepsilon'$
  \end{tabular}
\end{table}

\subsubsection{Towards a Unifying Framework}
The last two games (Table~\ref{table:probabilisticBisimGameIntroFKP}
from~\cite{FijalkowKP17} and
Table~\ref{table:probabilisticBisimMetricGameIntro} that seems new)
motivate a general framework that embraces both. There are some clear
analogies: the games are about \emph{indistinguishability} of states
$x,y$ under a class of \emph{observations} ($Z$ and $f$
respectively), and the \emph{predicates} usable in those observations
are subject to certain preservation properties
(bisimilarity-closedness in the former, and non-expansiveness in the
latter).

\subsection{A Codensity-Based Framework for Bisimilarity and Games}
\label{subsec:introContrib}
The main contribution of the current paper is a categorical framework
that derives a variety of bisimilarity notions and corresponding game
notions. The correspondence is proved once and for all on the categorical
level of generality. It covers the three examples introduced earlier
in~\S\ref{subsec:introBisimBisimGame}, much like the recent
categorical framework in~\cite{KoenigM18} does. However, our
fibration-based formalization has another dimension of generality. For
example, besides relations and metrics, our examples include what we
call \emph{bisimulation topology}.

\begin{figure}[tbp]
  \centering \scalebox{.8}{\begin{tikzpicture}[auto,
      semithick,remember picture, block/.style={rectangle, draw,
        minimum width=5em, text centered, rounded corners, minimum
        height=3em,text width=7em}, nrblock/.style={rectangle, draw,
        minimum width=5em, text centered, rounded corners, minimum
        height=3em,text width=5em} ]
      \draw[color=lightgray,very thick,dashed] (2.7,1) -- (2.7,-5);
      \node[block](n11){\textcircled{\small 1} codensity bisimilarity
        game}; \node[nrblock,right= 5em of
      n11](n12){\textcircled{\small 4} bisimilarity game};
      \node[block,below= 3em of
      n11](n21){\textcircled{\small 2} codensity bisimilarity\\
        $\nu\Phi^{\bOmega,\tau}\in \EE_{X}$}; \node[nrblock,right= 5em
      of n21](n22){\textcircled{\small 5} bisimilarity};
      \node[block,below= 2em of n21](n31){\textcircled{\small 3}
        codensity
        lifting~(\S{}\ref{sec:codensityBisimilarity},~\cite{KatsumataS15})
        $F^{\bOmega,\tau}\colon \EE\to \EE$}; \node[gray,above= 0.2em
      of n11] () {\Large categorical}; \node[gray,above= 0.2em of n12]
      () {\Large concrete}; \coordinate[left= 3em of n11](empty1);
      \coordinate[left= 3em of n31](empty3); \coordinate[right= 3em of
      n11](empty11); \coordinate[right= 1em of n12](empty12);
      \coordinate[right= 3em of n31](empty31); \coordinate[right= of
      n31](n32); \coordinate[right= 1em of n32](empty32);
      \draw[->] (n11) -- node {instantiates} (n12); \draw[->] (n21) --
      node {instantiates} (n22); \draw[->] (n11) -- node
      {\parbox{10em}{characterizes
          (Cor.~\ref{cor:soundnessAndCompletenessOfUntrimmedCodensityGame},~\ref{cor:soundnessAndCompletenessOfCodensityGame})}
      } (n21); \draw[->] (n31) -- node [auto=right] {induces
        (\S{}\ref{sec:codensityBisimilarity}, \cite{SprungerKDH18}) }
      (n21); \draw[->] (n31) -- (empty3) -- node
      [anchor=center,fill=white] {
        \parbox{4em}{induces
          (\S\ref{sec:untrimmedGame}--\ref{sec:trimmedGames})}
      } (empty1) -- (n11); \draw[->,gray,dotted,thick] (n22) edge[bend
      left=30] (n31);
    \end{tikzpicture}
  }
  \caption{Our Codensity-Based Framework for Bisimilarity and Games}
  \label{fig:codensityNotionsIntro}
\end{figure}
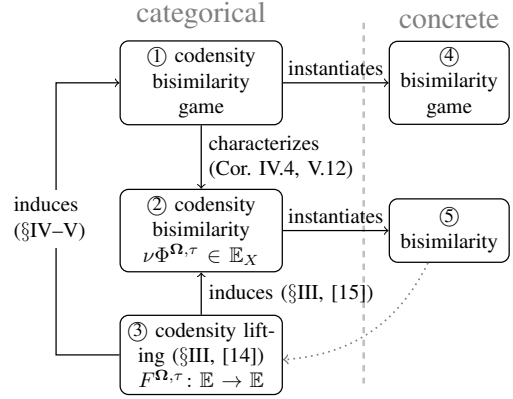

The overview of our categorical framework is in the left half of
Fig.~\ref{fig:codensityNotionsIntro}. We build on our
previous
works~\cite{KatsumataS15} and~\cite{SprungerKDH18}. In~\cite{KatsumataS15} a
general construction called \emph{codensity lifting} is introduced:
given a fibration $\fibp$ and parameters $(\bOmega,\tau)$ that embody
the kind of \emph{observations} we can make, a functor
$F\colon \CC\to\CC$ is lifted to $F^{\bOmega,\tau}\colon \EE\to\EE$.
In~\cite{SprungerKDH18}, codensity lifting is
used to introduce a generic family of bisimulation notions called
\emph{codensity bisimilarity}---see \textcircled{\small 2}. In this
paper, we extend these previous results by
\begin{itemize}
\item introducing the notion of \emph{codensity bisimilarity game}
  (\textcircled{\small 1}) that comes in two variants
  (\emph{untrimmed}~(\S\ref{sec:untrimmedGame}) and
  \emph{trimmed}~(\S\ref{sec:trimmedGames})),
\item establishing the correspondence between codensity bisimulations
  (\textcircled{\small 2}) and games (\textcircled{\small 1}) on a
  fibrational level of generality, and
\item working out several concrete examples (\textcircled{\small 4},
  \textcircled{\small 5}).
\end{itemize}

In general, devising a game notion (\textcircled{\small 4}) directly
from a bisimilarity notion (\textcircled{\small 5}) is far from
trivial. Indeed, doing so for an individual bisimilarity notion
has itself been deemed a scientific
novelty~\cite{DesharnaisLT08,FijalkowKP17}.
Our codensity-based framework (in the left half of
Fig.~\ref{fig:codensityNotionsIntro}) can automate \emph{part of} this
process in the following precise sense.

We derive concrete notions of bisimilarity
(\textcircled{\small 5}) and bisimilarity game (\textcircled{\small 4}) as
instances; then the correspondence between the two is guaranteed by
the categorical general result between \textcircled{\small 1} and
\textcircled{\small 2}.

We note, however, that this is no panacea. When one starts with a
given concrete notion of bisimilarity (\textcircled{\small 5}), their next
task would be to identify the right choice of the parameters
$\EE\xrightarrow{p}\CC,\bOmega,\tau$ for the codensity lifting
(\textcircled{\small 3}). This task is not easy in general:
we needed to get our hands dirty working out the examples in this
paper, \cite{KatsumataS15}, and~\cite{SprungerKDH18}. Nevertheless, we believe
that the required passage from \textcircled{\small 5} to
\textcircled{\small 3} is much easier than the direct derivation from
\textcircled{\small 5} to \textcircled{\small 4}, with our categorical framework providing templates of bisimilarity games (see Tables~\ref{table:untrimmedCodensityGame},~\ref{table:codensityBisimGame} and~\ref{table:multCodensityBisimGame}).
After all, our
framework identifies which part of the path from
\textcircled{\small 5} to \textcircled{\small 4} can be automated, and
which part remains to be done individually. This is much like what
many other categorical frameworks offer, as meta-level theories.

As an additional benefit, our categorical framework can be used to
\emph{discover} new bisimilarity notions (\textcircled{\small 5}),
starting from (choices of parameters for) \textcircled{\small 3}. We
believe those derived new bisimilarity notions are useful, since our
categorical theory embodies sound intuitions about observation,
predicate transformation, and indistinguishability---see
e.g.~\S\ref{subsec:fibrationPrelim}.

\subsection{Contributions}
\label{subsec:contribSummary}
Our main technical contributions are as follows.
\begin{itemize}
\item We introduce a categorical framework that uniformly describes
  various bisimulation notions (including metrics, preorders and
  topologies) and the corresponding game notions
  (Fig.~\ref{fig:codensityNotionsIntro}). The framework is based on
  coalgebras, fibrations, and codensity
  liftings in particular~\cite{KatsumataS15}. Our general game notion comes in two
  variants.
  \begin{itemize}
  \item The first (the \emph{untrimmed} codensity
    game:~\S{}\ref{sec:untrimmedGame}) arises naturally in a
    fibration, using its objects and arrows as possible moves.  The
    untrimmed game is theoretically clean, but it tends to have a huge
    arena.
  \item We therefore introduce a method that restricts these arenas,
    leading to the (\emph{trimmed}) codensity bisimilarity game
    (\S{}\ref{sec:trimmedGames}). The reduction method is also
    described in general fibrational terms, specifically using fibered
    separators and generating sets.
  \end{itemize}
\item From the general framework, we derive several concrete examples
  of bisimilarity and its related notions (\textcircled{\small 4} and
  \textcircled{\small 5} in
  Fig.~\ref{fig:codensityNotionsIntro}). They are listed in
  Table~\ref{table:codensityLiftings}.  Among them, a few bisimilarity
  notions seem new (especially the \emph{bisimulation topology}
  examples), and several game notions also seem new.
\item We discuss the \emph{transfer of codensity bisimilarity} by suitable fibered functors (\S{}\ref{sec:relatingDifferentSit}). As an example usage, we give an abstract proof of the
  fact that (usual) bisimilarity for Kripke frames  is necessarily an equivalence
  (Example~\ref{example:convBisimTransfer}).
\item
  Additionally, 
  we conduct investigations of the game notion
  in~\cite{FijalkowKP17}
  (Table~\ref{table:probabilisticBisimGameIntroFKP}) in concrete,
  non-categorical terms. For one, we obtain its variation for bisimulation
  metric (as we showed in
  Table~\ref{table:probabilisticBisimMetricGameIntro}).  We also give
  a direct proof of the equivalence to another game notion for probabilistic bisimilarity, previously
  introduced in~\cite{DesharnaisLT08}, by exhibiting a mutual translation of
  winning strategies (Appendix~\ref{appendix:gameEquivDirectProofProbBisim}).
\end{itemize}

\subsection{Related Work}\label{subsec:relatedWork}

Besides the one in~\cite{FijalkowKP17}, another game characterization
of probabilistic bisimulation has been given
in~\cite{DesharnaisLT08}. It is described later
in~\S{}\ref{sec:prelim}
(Table~\ref{table:probabilisticBisimGameDesharnais}). The latter game
has a bigger arena than the one in~\cite{FijalkowKP17}:
in~\cite{DesharnaisLT08} both players have to play a subset
$Z\subseteq X$, while in~\cite{FijalkowKP17} only Spoiler does so.

The work that is the closest to ours is the recent
work~\cite{KoenigM18} that studies bisimilarity games in a categorical
setting. Their formalization uses (co)algebras (following the
(co)algebraic generalization of the Kantorovich metric introduced
in~\cite{BaldanBKK18}), and therefore embraces a variety of different
branching types. The major differences between the two works are as
follows.
\begin{itemize}
\item Our current work is fibration-based (in particular
  $\CLatw$-fibrations), while~\cite{KoenigM18} is not. As a
  consequence, ours accommodates an additional dimension of generality
  by changing fibrations, which correspond to different indistinguishability
  notions (relation, metric, topology, preorder, measurable
  structures, etc.). In contrast, the
  works~\cite{KoenigM18} and~\cite{BaldanBKK18} deal exclusively with two
  settings: binary relations and pseudometrics.
\item A relationship to \emph{modal logic} is beautifully established
  in~\cite{KoenigM18}, while it is not done in this work. We expect
  our fibrational framework can accommodate modal logic too:
  fibrations have been used for categorical modeling of
  logics~\cite{Jacobs99a}. We leave this aspect to future work.

\item The categorical generalization~\cite{KoenigM18} is based on the
  game notion in~\cite{DesharnaisLT08}, while ours is based on that
  in~\cite{FijalkowKP17}. Therefore, for some bisimulation notions
  (including the bisimulation metric), we obtain a game notion with a
  smaller arena. Compare
  Table~\ref{table:probabilisticBisimMetricGameIntro} (an instance of
  ours) and Table~\ref{table:probabilisticBisimMetricGameKoenig} (an
  instance of~\cite{KoenigM18}).
\end{itemize}

There are a number of categorical studies of bisimilarity notions;
notable mentions include open map-based approaches~\cite{JoyalNW96} and
coalgebraic ones~\cite{Rutten00a,Jacobs16coalgBook}. The fibrational
approach we adopt also uses coalgebras; it was initiated
in~\cite{HermidaJ98} and pursued, e.g.,\
in~\cite{BonchiKP18,HasuoKC18,BonchiPPR14}, and~\cite{SprungerKDH18}. For example,
in the recent work~\cite{BonchiKP18}, fibrational generality is
exploited to study up-to techniques for bisimilarity metric. They use
the \emph{Wasserstein lifting} of functors introduced in~\cite{BaldanBKK18} instead of the codensity lifting that we use
(it generalizes the \emph{Kantorovich lifting} in~\cite{BaldanBKK18}, see Example~\ref{example:KantorovichLiftingSubsumed}). It is known~\cite{BaldanBKK18} that
the Wasserstein and Kantorovich liftings can differ in  general, while they coincide for some specific functors
such as the distribution functor.

Some of our new examples are topological: we derive what we call
\emph{bisimulation topology} and a game notion that characterizes
it. The relation between these notions and the existing works on
bisimulation and topology (including~\cite{BreugelMOW03,CuijpersR04})
is left as future work.

\subsection{Organization}
In~\S{}\ref{sec:prelim}, we present preliminaries on a general theory
of games (we can restrict to \emph{safety} games), and on
fibrations. For the latter, we focus on a class called
\emph{$\CLatw$-fibrations}, and argue that they offer
an appropriate
categorical abstraction of sets equipped with indistinguishability structures.
In~\S{}\ref{sec:codensityBisimilarity}, we present codensity lifting
and codensity bisimilarity (\textcircled{\small 2},
\textcircled{\small 3} in Fig.~\ref{fig:codensityNotionsIntro}). The
material is based on~\cite{SprungerKDH18}, but we introduce some
auxiliary notions needed for the correspondence with games. Our first
game notion (the \emph{untrimmed} one) is introduced
in~\S{}\ref{sec:untrimmedGame}; in~\S{}\ref{sec:trimmedGames}, we cut
down the arenas and obtain \emph{trimmed} codensity bisimilarity
game. The theory is further extended
in~\S{}\ref{sec:multipleObservationDomains}--\ref{sec:relatingDifferentSit}:
in~\S{}\ref{sec:multipleObservationDomains} we accommodate multiple
observation domains, and in~\S{}\ref{sec:relatingDifferentSit} we
discuss the transfer of codensity bisimilarities by full-faithful fibered functors preserving meets. These
categorical observations give rise to the concrete examples
in~\S{}\ref{sec:examples}.

Some proofs and details are deferred to Appendix~\ref{appendix:gameEquivDirectProofProbBisim}.

\section{Preliminaries}\label{sec:prelim}
We write $\pow\colon\Set\to\Set$ for the covariant powerset functor,
and $2$ for the two-point set $2=\{\bot,\top\}$.  We define the
function $\diamond:\pow 2\arrow 2$ called \emph{may-modality} by
$\diamond S=\top$ if and only if $\top\in S$. We write $\Eq_I$ for the
diagonal (equality) relation over a set $I$.

\subsection{Safety Games}\label{subsec:prelimGame}
Here we recall some standard game-theoretic notions and results. In
capturing bisimilarity-like notions, we can restrict ourselves to \emph{safety
  games}---they have a simple winning condition where every infinite
play is won by the same player (namely Duplicator). This winning
condition reflects the characterization of bisimilarity-like notions
by suitable \emph{greatest} fixed points; the correspondence
generalizes, for example, to the one between parity games and nested
alternating fixed points---see~\cite{Wilke01}. The term ``safety game''
occurs, e.g.,\ in~\cite{EhlersM12SYNT,BeyeneCPR14}.

Safety games are played between two players; in this paper, they are
called \emph{Duplicator} (D) and \emph{Spoiler} (S).  We restrict to
those games in which Duplicator and Spoiler alternate turns.

\begin{mydefinition}[safety game]\label{def:safetyGame}
  A \emph{(safety game) arena} is a triple
  $\mathcal{G}=(Q_{\Duplicator},Q_{\Spoiler},E)$ of a set
  $Q_{\Duplicator}$ of \emph{Duplicator's positions}, a set
  $Q_{\Spoiler}$ of \emph{Spoiler's positions}, and a \emph{transition
    relation}
  $E\subseteq (Q_{\Duplicator}\times
  Q_{\Spoiler})\cup(Q_{\Spoiler}\times Q_{\Duplicator})$. Hence
  $\mathcal{G}$ is a bipartite graph.  We require that
  $Q_{\Duplicator}$ and $Q_{\Spoiler}$ are disjoint, and that
  $Q_{\Duplicator}\cup Q_{\Spoiler}\neq \emptyset$. We write
  $Q=Q_{\Duplicator}\cup Q_{\Spoiler}$.

  For a position $q\in Q$, the elements of the set
  $\{q'\in Q\mid (q,q')\in E\}$ are called the \emph{possible moves}
  at $q$. Unlike some works, we allow positions that have no possible
  moves at them.

  A \emph{play} in an arena
  $\mathcal{G}=(Q_{\Duplicator},Q_{\Spoiler},E)$ is a (finite or
  infinite) sequence of positions $q_{0}q_{1}\dotsc$, such that
  $(q_{i-1},q_{i})\in E$ so long as $q_{i}$ belongs to the sequence.

  A play in $\mathcal{G}$ is \emph{won} by either player, according to
  the following conditions: 1) a finite play $q_{0}\dotsc q_{n}$ is
  won by Spoiler (or by Duplicator) if $q_{n}\in Q_{\Duplicator}$ (or
  $q_{n}\in Q_{\Spoiler}$ respectively); and 2) every infinite play
  $q_{0}q_{1}\dotsc$ is won by Duplicator.
\end{mydefinition}

\begin{mydefinition}[strategy, winning position]\label{def:strategy,winning position} In an arena  $\mathcal{G}=(Q_{\Duplicator},Q_{\Spoiler},E)$,
  a \emph{strategy} of Duplicator is a partial function
  $\sigma_{\Duplicator}\colon Q^{*}\times
  Q_{\Duplicator}\rightharpoonup Q_{\Spoiler}$; we require that
  $\sigma_{\Duplicator}(\vec{q}q)=q'$ implies $(q,q')\in E$. A
  \emph{strategy} of Spoiler is defined similarly, as a partial
  function
  $\sigma_{\Spoiler}\colon Q^{*}\times Q_{\Spoiler}\rightharpoonup
  Q_{\Duplicator}$ that returns a possible move at the last position
  in the history.

  Given an initial position $q\in Q$ and two strategies
  $\sigma_{\Duplicator}$ and $\sigma_{\Spoiler}$ for Duplicator and
  Spoiler respectively, the \emph{play} from $q$ induced by
  $(\sigma_{\Duplicator},\sigma_{\Spoiler})$ is defined in a natural
  inductive manner. The induced play is denoted by
  $\pi^{\sigma_{\Duplicator},\sigma_{\Spoiler}}(q)$.

  A position $q\in Q$ is said to be \emph{winning} for Duplicator if
  there exists a strategy $\sigma_{\Duplicator}$ of Duplicator such
  that, for any strategy $\sigma_{\Spoiler}$ of Spoiler, the induced
  play $\pi^{\sigma_{\Duplicator},\sigma_{\Spoiler}}(q)$ is won by
  Duplicator.

  In what follows, for simplicity, we restrict the initial position
  $q$ of a play $\pi^{\sigma_{\Duplicator},\sigma_{\Spoiler}}(q)$ to
  be in $Q_{\Spoiler}$. (Note that Spoiler's position can be winning
  for Duplicator.)
\end{mydefinition}

Winning positions of safety games are witnessed by \emph{invariants}
(Prop.~\ref{prop:invariantsAndWinningPositions}), which is a
well-known fact.
\begin{mydefinition}[invariant]\label{def:invariant}
  Let $\mathcal{G}=(Q_{\Duplicator},Q_{\Spoiler},E)$ be an arena. A
  subset $P\subseteq Q_{\Spoiler}$ is called an \emph{invariant} for
  Duplicator if, for each $q\in P$ and any possible move
  $q'\in Q_{\Duplicator}$ at $q$, there exists a possible move $q''$
  at $q'$ that is in $P$. That is,
  \begin{math}
    \forall q\in P.\, \forall q'\in Q_{\Duplicator}.\,
    \bigl( \,      (q,q')\in E
    \;\Rightarrow\;
    \exists q''\in Q_{\Spoiler}.\,
    \, (q',q'')\in E \land q''\in P
    \,\bigr).
  \end{math}
\end{mydefinition}

\begin{myproposition}\label{prop:invariantsAndWinningPositions}
  \begin{enumerate}
  \item Any position $q\in P$ in an invariant $P$ for Duplicator is
    winning for Duplicator.
  \item Invariants are closed under arbitrary union. Therefore, there
    exists a largest invariant for Duplicator.
  \item The largest invariant for Duplicator coincides with the set of
    winning positions for Duplicator in $Q_{\Spoiler}$. \myqed
  \end{enumerate}
\end{myproposition}

Examples of safety games have been given in
Tables~\ref{table:probabilisticBisimGameIntroFKP}--\ref{table:probabilisticBisimMetricGameIntro}. We
present two other examples
(Tables~\ref{table:probabilisticBisimGameDesharnais}--\ref{table:probabilisticBisimMetricGameKoenig}).

\begin{myexample}[alternative games for probabilistic bisimilarity and bisimulation metric]\label{example:gamesByDesharnaisAndKoenig}
  In~\cite{DesharnaisLT08}, a game notion that characterizes
  (qualitative) probabilistic bisimilarity is presented. It is in
  Table~\ref{table:probabilisticBisimGameDesharnais}, presented in a
  slightly adapted form.

  This game notion is categorically generalized in~\cite{KoenigM18};
  the generalization has freedom in the choice of coalgebra functors
  (i.e.\ branching types), as well as in the choice between relations
  and metrics. The instance of this general game notion for
  bisimulation metric is shown in
  Table~\ref{table:probabilisticBisimMetricGameKoenig}.

  The two games
  (Tables~\ref{table:probabilisticBisimGameDesharnais}--\ref{table:probabilisticBisimMetricGameKoenig})
  characterize the same bisimilarity-like notions as the games in
  Tables~\ref{table:probabilisticBisimGameIntroFKP}--\ref{table:probabilisticBisimMetricGameIntro}, respectively;
  so they are equivalent. We can go further and give a
  direct equivalence proof by mutually translating winning strategies.
  Such a proof is not totally trivial; we do so for the pair for
  probabilistic bisimilarity. 
  See Appendix~\ref{appendix:gameEquivDirectProofProbBisim}.

  We note that the game in
  Table~\ref{table:probabilisticBisimMetricGameIntro} (an instance of
  our current framework) is simpler than Table~\ref{table:probabilisticBisimMetricGameKoenig} (an
  instance of~\cite{KoenigM18}). Table~\ref{table:probabilisticBisimMetricGameIntro} is not only structurally simpler (it has fewer rows), but its set of moves are smaller too, asking for functions $X\to [0,1]$ only at one place.

  \begin{table}[htbp]
    \caption{The Game for Probabilistic Bisimilarity,
      from~\cite{DesharnaisLT08}}
    We omit labels that are needed to distinguish $(x,y)\in X^{2}$ (an
    S-position) from $(s,t)\in X^{2}$ (a D-position).
    \label{table:probabilisticBisimGameDesharnais}
    \centering \renewcommand{\arraystretch}{1.2}
    \begin{tabular}{l|l|l}
      position & pl.\ &  possible moves    \\\hline\hline
      $(x,y)\in X^{2}$
               & S &
                     $(s,t)\in X^{2}$ s.t.\ $\{s,t\}=\{x,y\}$
      \\\hline
      $(s,t)\in X^{2}$
               & D &
                     $(Z,Z')$ s.t.\ $Z\subseteq Z'\subseteq X$
      \\
               & & and $c(s)(Z)\le c(t)(Z')$ \\\hline
      $(Z,Z')\in (\pow X)^{2}$
               &
                 S
                      &
                        $(Z,y')\in \pow X\times X$ s.t.\  $y'\in Z'\setminus Z$
      \\\hline
      $(Z,y')\in \pow X\times X$
               &
                 D
                      &
                        $(x',y')\in X^2$ s.t.\ $x'\in Z$
    \end{tabular}

    \vspace{2em}
    \caption{The Game for Bisimulation Metric, from~\cite{KoenigM18}}
    \label{table:probabilisticBisimMetricGameKoenig}
    \centering \renewcommand{\arraystretch}{1}
    \begin{tabular}{l|l|l}
      position & pl.\ &  possible moves    \\\hline\hline
      $(x,y,\varepsilon)\in X^{2}\times[0,1]$
               & S &
                     $(s,t)\in X^{2}$ s.t.\ $\{s,t\}=\{x,y\}$, \\
               &&
                  and
                  $f\colon X\to [0,1]$
      \\\hline
      $(s,t,f,\varepsilon) \in$
               & D &
                     $g\colon X\to [0,1]$ such that \\[+.3em] $X^2 \times {[0,1]}^X \times
      [0,1]$ && $\max \{0, \expect{c(s)}{f} - \expect{c(t)}{g}\} \leq \varepsilon$
      \\\hline
      $(f, g, \varepsilon) \in ({[0,1]}^X)^2$ & S & $(i, j) \in {\left({[0,1]}^X\right)}^2$
                                                    such that
      \\
               &&$\{i,j\} = \{f,g\}$, and $x^\prime \in X$
      \\\hline
      $(x^\prime, i, j, \varepsilon) \in$ & D & $(x^\prime, y^\prime,
                                                \varepsilon^\prime) \in X^2 \times [0,1]$ such that
      \\
      $X \times ({[0,1]}^X)^2 \times
      [0,1]$ &&
                $i(x^\prime) \leq j(y^\prime)$, and\\
               & & $\varepsilon^\prime =             j(y^\prime)-i(x^\prime)$
    \end{tabular}	 \end{table}
\end{myexample}

Our categorical framework based on codensity liftings (presented in
later sections) covers
Tables~\ref{table:probabilisticBisimGameIntroFKP}--\ref{table:probabilisticBisimMetricGameIntro}
but not
Tables~\ref{table:probabilisticBisimGameDesharnais}--\ref{table:probabilisticBisimMetricGameKoenig}. Accommodation
of the latter two is future work.

\subsection{
  $\CLatw$-fibrations
}\label{subsec:fibrationPrelim}

\subsubsection{Definition and Properties}
Here we sketch a basic theory of fibrations---see, e.g.,~\cite{Jacobs99a}
for a comprehensive account. In particular, we focus on a class of
poset fibrations called \emph{$\CLatw$-fibrations}. We observe that
the simple axiomatics of the class adequately capture all the examples
of interest---and hence the mathematical essences of the logical
phenomena that we wish to model.

Our exposition here is largely based on that
in~\cite{SprungerKDH18}. However, in this paper we introduce new
notation and terminology (such as \emph{indistinguishability order}
and \emph{decent map})---see~\S{}\ref{subsubsec:CLatwFibNotationAndIntuition}. They help to
further clarify the intuitions.

A formal definition is as follows.
                                         (See Appendix~\ref{appendix:CLatwFib} for a rather gentle introduction to $\CLatw$-fibration.)
\begin{mydefinition}[$\CLatw$-fibration]\label{def:CLatwFib}
  A \emph{$\CLatw$-fibration} is a fibration $\fibp$ such that each
  fiber $\EE_{X}$ (for each $X\in \CC$) is a complete lattice, and
  each pullback functor $f^{*}\colon\EE_{Y} \to\EE_{X}$ (for each
  $f\colon X\to Y$ in $\CC$) preserves all meets $\bigsqcap$.
\end{mydefinition}
Via the Grothendieck construction, a $\CLatw$-fibration is in a
bijective correspondence with a functor
$F_\EE\colon\CC^{\op}\to \CLatw$, where $\CLatw$ is the category of
complete lattices and functions preserving all meets---see~\cite{Jacobs99a}
and~\cite{HasuoKC18},
as well 
                                         as~Appendix~\ref{appendix:CLatwFib}.
The functor $F_\EE$ assigns
\begin{itemize}
\item a complete lattice $\EE_{X}$ (called the \emph{fiber} over $X$)
  to each $X\in\CC$, and
\item a function $f^{*}\colon \EE_{Y}\to \EE_{X}$ preserving all meets
  to each $f\colon X\to Y$ in $\CC$. The map $f^{*}$ is called a
  \emph{pullback}; it is also called a \emph{pullback functor} since,
  in the general theory of fibrations, a fiber $\EE_{X}$ is a category
  rather than a poset.
\end{itemize}
Although the \emph{indexed category} presentation
$F_\EE\colon\CC^{\op}\to \CLatw$ may be more intuitive at first, we
shall stick to the \emph{fibration} presentation $\fibp$ since we will
eventually need some global structures in the \emph{total category}
$\EE$. It turns out that $\CLatw$-fibrations are special kinds of
{\em{topological functor}} {\cite{HERRLICH1974125}} such that each
fiber category is a poset.  Topological functors are a well-studied
topic, and many examples and results are available; a good summary is
found in {\cite{Adamek04abstractand}}.

The use of poset fibrations is common in categorical modeling of
logics~\cite{HasuoKC18,BonchiPPR14}. $\CLatw$-fibrations additionally
require fibered small meets; this simple assumption turns out to be a
mathematically powerful one.
\begin{myproposition}\label{prop:propertiesOfCLatwFib}
  Let $\fibp$ be a $\CLatw$-fibration.
  \begin{enumerate}
  \item $p$ is split, and faithful as a functor.
  \item Each arrow $f\colon X\to Y$ has its \emph{pushforward}
    $f_{*}\colon \EE_{X}\to \EE_{Y}$, so that an adjunction
    $f_*\dashv f^*$
    is formed. This is a consequence of Freyd's adjoint functor
    theorem; it makes $p$ a \emph{bifibration}~\cite{Jacobs99a}.
  \item $p^{\op}\colon \EE^{\op}\to\CC^{\op}$ is also a
    $\CLatw$-fibration.
  \item The change-of-base \cite[Lemma 1.5.1]{Jacobs99a} of $p$ along
    any functor $H:\DD\arrow\CC$ is also a $\CLatw$-fibration.
  \item If $\CC$ is (co)complete, then the total category $\EE$ is
    also (co)complete. This follows
    from~\cite[Prop.~9.2.1]{Jacobs99a}.\myqed
  \end{enumerate}
\end{myproposition}

\subsubsection{Notation, Terminology and Intuitions}
\label{subsubsec:CLatwFibNotationAndIntuition}
Our view of a $\CLatw$-fibration $\fibp$ is that it equips objects of
$\CC$ with what we call \emph{indistinguishability structures}. This suits our
purpose, since various bisimilarity-like notions are all about degrees
of indistinguishability between (the behaviors of) states of a
system. We present examples later
in~\S{}\ref{subsubsec:CLatwFibexamples}.
\begin{mynotation}[indistinguishability predicate/order]
  Let $\fibp$ be a $\CLatw$-fibration.  An object $P\in \EE_{X}$
  in the fiber category $\EE_{X}$ (i.e.\ an element of the complete lattice $\EE_{X}$)
  is called an \emph{indistinguishability predicate} over $X$. Our
  view is that $P$ is an additional structure on $X$; therefore, as a
  convention, an object $P\in \EE_{X}$ shall also be denoted by
  $(X,P)\in \EE_{X}$.

  Each fiber $\EE_{X}$ is a complete lattice; its order is denoted by
  $\sqsubseteq$ and called the \emph{indistinguishability order} over
  $X$. Intuitively, $P\sqsubseteq Q$ means that $Q$ has a greater
  degree of indistinguishability than $P$---that is, $Q$ is coarser
  than $P$, and $P$ is more discriminating than $Q$.

  The supremum and infimum with respect to the indistinguishability
  order $\sqsubseteq$ are denoted by $\bigsqcup$ and $\bigsqcap$
  respectively.
\end{mynotation}

\begin{mydefinition}[decent map]\label{def:decentMap}
  Let $\fibp$ be a $\CLatw$-fibration, $f\colon X\to Y$ be an arrow in
  $\CC$, $(X,P)\in \EE_{X}$ and $(Y,Q)\in \EE_{Y}$ be objects in
  the fibers. We say that $f$ is \emph{decent (from $P$ to $Q$)} if
  there exists a necessarily unique arrow $\dot {f}\colon P\arrow Q$
  in $\EE$ such that $p\dot f=f$. We write $f\colon (X,P)\darrow(Y,Q)$ in this case.
  The following equivalences follow.
  \begin{displaymath}
    f:(X,P)\darrow (Y,Q)\iff
    P\sqsubseteq f^*Q\iff
    f_*P\sqsubseteq Q
  \end{displaymath}
  We write
  $f\colon (X,P)\ndarrow (Y,Q)$ if $f$ is {\em not} decent.
\end{mydefinition}
The notion of decency is a fibered generalization of
continuity, non-expansiveness, relation-preservation, etc. Decency $f\colon (X,P)\darrow (Y,Q)$ means $f$
\emph{respects} indistinguishability, carrying $P$-indistinguishable
elements
to $Q$-indistinguishable ones.

\subsubsection{Examples}
\label{subsubsec:CLatwFibexamples}

\begin{table*}[htbp]
  \caption{$\CLatw$-Fibrations}
  \label{table:fibex}
  \begin{tabular}{l|l|l|l|l} fibration & indistinguishability
                                         structure & decent map & $P\sqsubseteq Q$ & $\bigsqcap P_i$
    \\
    \hline\hline $\Top\to\Set$ & topology & continuous func. &
                                                               $P
                                                               \supseteq
                                                               Q$ &
                                                                    generated
                                                                    from
                                                                    $\bigcup
                                                                    P_i$
    \\ \hline $\Meas\to\Set$ & $\sigma$-field & measurable func. &
                                                                   $P\supseteq
                                                                   Q$
                                                                                   & generated from
                                                                                     $\bigcup P_i$ \\
    \hline $\PMet\to\Set$ & pseudometric & non-expansive func. &
                                                                 $\forall x,y.\, P(x,y)\ge
                                                                 Q(x,y)$
                                                                                   & $(x,y)\mapsto \sup_{i} P_i(x,y)$ \\
    \hline $\ERel\to\Set$ & endorelation & relation preserving func. &
                                                                       $P\subseteq
                                                                       Q$ & $\bigcap P_i$ \\
    \hline $\Pre\to\Set$ & preorder & monotone func. & $P\subseteq Q$
                                                                                   & $\bigcap P_i$\\ \hline
    $\EqRel\to\Set$ & equivalence relation
                                                   & relation preserving func. & $P\subseteq Q$ &  $\bigcap P_i$\\

  \end{tabular}
  \centering

  \vspace{1em}
  \caption{Codensity Lifting of Functors}
  \label{table:codensityLiftings}
  \begin{tabular}{r|l|l|l|l|l}
    &fibration $\fib{\EE}{p}{\CC}$ & functor $F\colon\CC\to\CC$ & obs.\ dom.\ $\bOmega$ & modality $\tau$ & lifting $F^{\bOmega,\tau}$ of $F$     \\\hline\hline
    \newtag{1}{row:lowerpreorder}&    $\Pre\to\Set$ & powerset $\pow $ & $(2,\le)$ & $\diamond\colon \pow 2\to2$ & lower preorder~\cite{KatsumataS15} \\\hline
    \newtag{2}{row:upperpreorder}&    $\Pre\to\Set$ & powerset $\pow $ & $(2,\ge)$ & $\diamond\colon \pow 2\to2$ & upper preorder~\cite{KatsumataS15} \\\hline
    \newtag{3}{row:ERelEq}&    $\ERel\to\Set$ & powerset $\pow $ & $(2,\Eq_2)$ & $\diamond\colon \pow 2\to 2$ & (for bisimulation, see Ex.~\ref{ex:erelpowlift} \&~\ref{example:convBisimTransfer}) \\\hline
    \newtag{4}{row:EqRelBisim}&    $\EqRel\to\Set$ & powerset $\pow$ & $(2,\Eq_2)$ & $\diamond\colon \pow 2\to2$ & (for bisimulation, see Ex.~\ref{ex:eqrelpowlift} \&~\ref{example:convBisimTransfer}) \\\hline
    \newtag{5}{row:Kantorovich}&    $\PMet\to\Set$ & subdistrib.\ $\sdist$ & $([0,1],d_{[0,1]})$ & $e\colon \sdist[0,1]\to[0,1]$ & Kantorovich metric
    \\\hline
    \newtag{6}{row:Hausdorff}&    $\PMet\to\Set$ & powerset $\pow $ & $([0,1],d_{[0,1]})$ & $\inf\colon \pow [0,1]\to[0,1]$ & Hausdorff pseudometric 
                                                                                                                                                                                                                                                             (cf.\ Appendix~\ref{subsec:hausdorff})
    \\\hline
    \newtag{7}{row:contiKantorovich}&    $U^*(\PMet)\to\Meas$ & sub-Giry $\mathcal{G}_{\le 1}$ & $([0,1],d_{[0,1]})$ & $e\colon \mathcal{G}_{\le 1}[0,1]\to[0,1]$ & Kantorovich metric
    \\\hline
    \newtag{8}{row:convexpreorder}${}^{\dagger}$&    $\Pre\to\Set$& powerset $\pow $ & $(2,\le),(2,\ge)$  & $\diamond\colon \pow 2\to2$ & convex preorder~\cite{KatsumataS15} \\\hline
    \newtag{9}{row:probBisim}${}^{\dagger}$&    $\EqRel\to\Set$ & subdistrib.\ $\sdist$ & $(2,\Eq_2)$   & $(\tau_r\colon
                                                                                                          \sdist2\to 2)_{r\in[0,1]}$  & (for prob.\ bisim., see \S{}\ref{subsec:FKPGame}) \\\hline
    \newtag{10}{row:bisimTop}${}^{\dagger}$&    $\Top\to\Set$& $2\times(\place)^\Sigma$ &Sierpinski space & (see Ex.~\ref{example:bisimTop}) &  (for bisim.\ topology, see Ex.~\ref{example:bisimTop})
  \end{tabular}
  \parbox{\textwidth}{The fibration $U^*(\PMet)\to\Meas$ is obtained as a change-of-base, pulling back $\PMet\to\Set$ along $U\colon \Meas\to \Set$.
    $d_{[0,1]}$ denotes the Euclidean metric on the unit interval $[0,1]$.
    The modality $\diamond$ is introduced in the beginning of
    \S\ref{sec:prelim}.
    The functions $e\colon \sdist[0,1]\to[0,1]$ and $e\colon \mathcal{G}_{\le 1}[0,1]\to[0,1]$ both return expected values.
    The lower, upper and convex preorders are known for powerdomains;
    see e.g.~\cite{TixKP05}.
    The function $\tau_r\colon \sdist 2\to2$ is defined by
    $\tau_r(p) =\top$ if $p(\top) \ge r$, and $\tau_r(p) =\bot$ otherwise.

    The  examples marked with $\dagger$ involve multiple modalities and observation domains. The extension that allows such is described later in~\S{}\ref{sec:multipleObservationDomains}.
  }
\end{table*}
As shown in \Cref{table:fibex}, various well-known categories can be
seen as categories that equip sets with certain indistinguishability
structures.  The evident forgetful functors from the total categories
($\Top$, $\Meas$, etc.) to $\Set$ in~\Cref{table:fibex} are all
$\CLatw$-fibrations.

Specifically, $\Top$ is the category of topological spaces and
continuous maps; $\Meas$ is that of measurable spaces and measurable
maps; $\PMet$ is that of $1$-bounded pseudometric spaces (where a
\emph{pseudo}-metric is a metric without the condition
$d(x,y)=0\Rightarrow x=y$) and non-expansive maps; $\ERel$ is that of
sets with endorelations $(X,R\subseteq X^{2})$ and relation-preserving
maps; $\Pre$ is that of preordered sets and monotone maps; and
$\EqRel$ is that of sets with equivalence relations and
relation-preserving maps---see~\cite{SprungerKDH18} for details.

Note that, in $\Top$ and $\Meas$, the indistinguishability order is
the opposite of the inclusion order. Therefore the meet of a family of
indistinguishability structures computed as the one generated
from the \emph{union} of the family.

Another class of examples is given as follows: for any well-powered
category $\BB$ admitting small limits, the subobject fibration of
$\BB$ is a $\CLatw$-fibration. All the algebraic categories over
$\Set$ and Grothendieck topoi fall into this class.  On the other
hand, the forgetful functors from algebraic categories over $\Set$ are
rarely ($\CLatw$-)fibrations.

\section{Codensity Bisimilarity}\label{sec:codensityBisimilarity}
We introduce \emph{codensity lifting} (\textcircled{\small 3} in
Fig.~\ref{fig:codensityNotionsIntro}) and \emph{codensity
  bisimilarity} (\textcircled{\small 2}) based on~\cite{SprungerKDH18}.
These turn out to subsume many bisimilarity-like notions in the
literature.  The material
in~\S{}\ref{subsec:codensityLifting}--\ref{subsec:codensityBisim} is
largely from~\cite{SprungerKDH18}; \S{}\ref{subsec:jointCodensityLifting} is new, paving the way to
codensity bisimilarity games presented in later sections.

\subsection{Codensity Lifting}\label{subsec:codensityLifting}

\begin{mydefinition}[codensity lifting $F^{\bOmega,\tau}$
  \cite{SprungerKDH18}]\label{def:codensityLifting}
  Let
  $\fibp$ be a $\CLatw$-fibration, and
  $F : \CC \rightarrow \CC$ be a functor. A {\em parameter of codensity
    lifting} of $F$ along $p$ is a pair
  of
  \begin{itemize}
  \item a $\CC$-arrow $\tau : F \Omega \rightarrow \Omega$ (i.e.\ an $F$-algebra) called
    \emph{modality}~\cite{Hasuo15CMCSJournVer,HinoKHJ16} and
  \item an $\EE$-object $\bOmega$ above $\Omega$ called
    \emph{observation domain}.
  \end{itemize}
  The \emph{codensity lifting} of $F\colon \CC\rightarrow\CC$ with parameter
  $(\bOmega,\tau)$ is the endofunctor $F^{\bOmega,\tau}:\EE\arrow\EE$ defined as follows.
  On objects,
  \begin{displaymath}
    F^{\bOmega,\tau} P =
    \bigsqcap_{k \in \EE (P, \bOmega)} \bigl(\tau \circ F\bigl(p (k)\bigr)\bigr)^{\ast} \bOmega.
  \end{displaymath}
  Its action on arrows is as follows.  It is not hard to see that, for
  each arrow $l:P\arrow Q$ in $\EE$, the arrow $F(p(l))$ is decent from
  $F^{\bOmega,\tau} P$ to $F^{\bOmega,\tau} Q$. Then we define
  $F^{\bOmega,\tau} l\colon F^{\bOmega,\tau} P\to F^{\bOmega,\tau} Q$  to be the unique arrow in $\EE$
  above $F(p(l))$.
\end{mydefinition}
An alternative description is possible. When $\EE$ has powers
$\pitchfork$ and $p$ preserves them, $F^{\bOmega,\tau}$ is
characterized as the following
pullback
in the fibration
$\fib{[\EE,\EE]}{[\EE,p]}{[\EE,\CC]}$:
\begin{displaymath}
  \xymatrix@R=1em@C+1em{
    [\EE,\EE] \ar[d]^-{[\EE,p]} & F^{\bOmega,\tau} \ar@{-->}[r] &  \EE(-,\bOmega)\pitchfork\bOmega \\
    [\EE,\CC] &    F\co p \ar[r]_-{\alpha}  & \EE(-,\bOmega)\pitchfork\Omega
  }
\end{displaymath}
where $\alpha_P=\langle\tau\circ F(p(k))\rangle_{k\in\EE(P,\bOmega)}$
is the tupling. A similar characterization of codensity
liftings of monads is in \cite{KatsumataS15}.

Table~\ref{table:codensityLiftings} lists concrete examples of codensity
liftings, with various fibrations $p$, functors $F$, and
parameters $(\bOmega,\tau)$. Some of
them
coincide with
known notions. For example, the entry~\ref{row:Kantorovich} of the table says that the functor $(\sdist)^{\bOmega,\tau}$, with the designated $\bOmega$ and $\tau$, carries a metric space $(X,d)$ to the set $\sdist X$ equipped with the well-known
Kantorovich metric $\mathcal{K}(d)$ induced by $d$---see~(\ref{eq:KantorovichLiftingIntro}).

Besides the functors listed in the table, there are some natural
ways to systematically lift polynomial functors, by defining
$\tau\colon F\Omega\to \Omega$ in an inductive manner---see,
e.g.,~\cite{BonchiKP18}.

\begin{myexample}\label{ex:eqrelpowlift}
  Let us closely look at the entry~\ref{row:EqRelBisim} of~\Cref{table:codensityLiftings}. 
  There we codensity-lift the
  covariant
  powerset functor $\pow$ along the $\CLatw$-fibration
  $\fib{\EqRel}{}{\Set}$. We use the parameter
  $((2,\Eq_2),\diamond)$, where
  $\diamond:\pow 2\arrow 2$ is the modality given in
  \S{}\ref{sec:prelim}.

  We shall abbreviate $(2,\Eq_{2})$ by $\Eq_{2}$---a notational convention that is used throughout the paper. 

  Then   $\pow^{\Eq_{2},\diamond}(X,R)$ relates $S,T\in\pow X$ if and only if
  \begin{align*}
    \begin{array}{l}
      \fa{k\colon X\to2}\bigl(\,(\fa{(x,y)\in R} k(x)= k(y))\bigr. \\
      \bigl.\implies\bigl((\ex{x\in S}k(x)=\top)\iff(\ex{x\in T}k(x)=\top)\bigr)\,\bigr).
    \end{array}  
  \end{align*}
  Straightforward calculation shows that this is equivalent to
  \begin{align*}\small
    \begin{array}{l}
      (\fa{x\in S}\ex{y\in T}(x,y)\in R )  \wedge
      (\fa{y\in T}\ex{x\in S}(x,y)\in R).
    \end{array}  
  \end{align*}
  This lifting is the restriction of the standard relational
  lifting of $\pow$ along $\fib{\ERel}{}{\Set}$,
  which is used for the usual bisimulation notion for Kripke frames,
  to $\EqRel$.
\end{myexample}

\begin{myexample}\label{ex:erelpowlift}
  In the entry~\ref{row:ERelEq} of~\Cref{table:codensityLiftings},  we codensity-lift $\pow$ along the $\CLatw$-fibration
  $\fib{\ERel}{}{\Set}$ (instead of $\fib{\EqRel}{}{\Set}$) with the parameter $\bigl((2,\Eq_2),\diamond\bigr)$.

  The characterization of $\pow^{\Eq_2,\diamond}(X,R)$ is
  slightly involved.  Its relation part relates $S,T\in\pow X$ if and only if
  \begin{align*}
    & (\fa{x\in S}\ex{y\in T}(x,y)\in R^{\mathrm{eq}} )~\wedge\\
    & (\fa{y\in T}\ex{x\in S}(x,y)\in R^{\mathrm{eq}}),
  \end{align*}
  where $R^{\mathrm{eq}}$ denotes the equivalence closure of $R$. 
\end{myexample}

It is not clear at this stage whether the codensity bisimilarities induced by the above liftings
(Examples~\ref{ex:eqrelpowlift}--\ref{ex:erelpowlift}, i.e.\ the entries~\ref{row:EqRelBisim} \&~\ref{row:ERelEq} of Table~\ref{table:codensityLiftings}) coincide with the usual bisimilarity notion for Kripke frames. This is because of 
the involvement of mandatory equivalence closures---specifically by the use of $\EqRel$ in Example~\ref{ex:eqrelpowlift}, and by the occurrence of $(\place)^{\mathrm{eq}}$ in Example~\ref{ex:erelpowlift}. 
Later, in~\Cref{example:convBisimTransfer}, we prove that both of the codensity bisimilarities indeed coincide with the usual bisimilarity notion. The proof relies crucially on  transfer of codensity liftings via fibered functors.

\begin{myexample}  \label{example:KantorovichLiftingSubsumed}
  Here we follow~\cite[Example 3]{SprungerKDH18} and show that
  codensity lifting generalizes the
  \emph{Kantorovich lifting} of functors introduced in~\cite{BaldanBKK18}.
  Take $\fib{\PMet}{}{\Set}$ as the $\CLatw$-fibration  $p$ in
  Def.~\ref{def:codensityLifting}.
  As $\bOmega$, we take $\Omega=[0,1]$ with the usual Euclidean metric $d_{[0,1]}$.
  There is freedom in the choice of a modality $\tau\colon F\Omega\to \Omega$---this corresponds to what is called an \emph{evaluation function} in~\cite{BaldanBKK18}.
  This
  way we recover the  Kantorovich lifting in~\cite{BaldanBKK18} as $F^{\bOmega,\tau}$.
\end{myexample}

\subsection{Codensity Bisimilarity}\label{subsec:codensityBisim}
In~\cite{SprungerKDH18}, \emph{codensity bisimulation} and
\emph{bisimilarity} are introduced. Recall that a
\emph{coalgebra} $c\colon X\to FX$ is a categorical presentation of
state-based transition systems, such as automata, Markov chains, etc.---see, e.g.,~\cite{Jacobs16coalgBook,Rutten00a}, and
also~\S{}\ref{sec:examples}.

\begin{mydefinition}[codensity
  bisimulation]\label{def:codensityBisimulation}
  Assume the setting of Def.~\ref{def:codensityLifting}.  Let
  $c:X\to FX$ be an $F$-coalgebra.  An object $P\in\EE_X$ is a {\em
    ($(\bOmega,\tau)$-)codensity bisimulation} over $c$ if
  $c:(X,P) \darrow (FX,F^{\bOmega,\tau} P)$; that is, $c$ is decent
  with respect to the designated indistinguishability structures on $X$ and $FX$.
\end{mydefinition}

We move on to the characterization of codensity bisimulations as (post-)fixed points of suitable predicate transformers.
\begin{mydefinition}[predicate transformer $\Phi^{\bOmega,\tau}$]\label{def:defOfPhi}
  Assume the setting of Def.~\ref{def:codensityBisimulation}.  We
  define a \emph{predicate transformer}
  $\Phi_c^{\bOmega,\tau} : \EE_X \rightarrow \EE_X$ with respect to $c$
  and $F^{\bOmega,\tau}$ by:
  \begin{equation}\label{eq:defOfPhi}
    \Phi_c^{\bOmega,\tau} P
    = c^{\ast} (F^{\bOmega,\tau} P), \text{ that is, } \bigsqcap_{k \in \EE
      (P, \bOmega)} \bigl(\tau \circ F (p(k)) \circ c\bigr)^{\ast} \bOmega.
  \end{equation}
\end{mydefinition}

\begin{mytheorem}\label{thm:codensityBisimChara}
  Assume the setting of Def.~\ref{def:codensityBisimulation}.  For any
  $P\in\EE_X$, the following are equivalent.
  \begin{enumerate}
  \item $c:(X,P)\darrow(FX, F^{\bOmega,\tau} P)$; that is, $P$ is a
    codensity bisimulation over $c$
    (Def.~\ref{def:codensityBisimulation}).
  \item $P\sqsubseteq\Phi_c^{\bOmega,\tau} P$.
  \item For each $k\in\CC(X,\Omega)$,
    $k:(X,P)\darrow (\Omega,\bOmega)$ implies
    $\tau \circ F k \circ c: (X,P)\darrow (\Omega,\bOmega)$.  \myqed
  \end{enumerate}
\end{mytheorem}

The predicate transformer $\Phi_c^{\bOmega,\tau}$ is a monotone map from
the complete lattice $\EE_{X}$ to itself. Therefore, by the
Knaster--Tarski theorem, the greatest post-fixed point of
$\Phi_c^{\bOmega,\tau}$ exists and it is the greatest fixed point of
$\Phi_c^{\bOmega,\tau}$.
\begin{mydefinition}[codensity bisimilarity $\nu \Phi_c^{\bOmega,\tau}$]
  \label{def:codensityBisimilarity}
  Assume the setting of Def.~\ref{def:codensityBisimulation}.  The
  greatest codensity bisimulation, whose existence is guaranteed by
  the above arguments,
  is called the \emph{codensity bisimilarity}. It is denoted by
  $\nu \Phi_c^{\bOmega,\tau}$.
\end{mydefinition}

Some bisimilarity notions, including bisimilarity of
deterministic automata
(\S{}\ref{subsec:DFA}), are accommodated in the
generalized framework with multiple observation domains---see~\S{}\ref{sec:multipleObservationDomains}.

\begin{myexample}[bisimulation metric]\label{example:behMetric}
  Consider the $\CLatw$-fibration $\fib{\PMet}{}{\Set}$ and the
  subdistribution functor $\sdist\colon\Set\to\Set$. Recall that 
  \begin{math}
    \textstyle
    \sdist(X)=\{p\colon X\to [0,1]\mid \sum_{x\in X}p(x)\le 1\}.
  \end{math}
  As a parameter of codensity lifting, we take 
  $(\bOmega,\tau)=\bigl(\,\bigl([0,1],d_{[0,1]}\bigr),\,e\colon \sdist [0,1]\to[0,1]\,\bigr)$,
  where $e$ is the {\em expectation function} 
  \begin{math}
    e(p)=\sum_{r\in[0,1]}r\cdot p(r)
  \end{math}
  and $d_{[0,1]}$ is
  the Euclidean metric.
  Let $c\colon X\to \sdist X$ be a coalgebra, identified with a Markov chain.

  The codensity bisimilarity in this setting coincides with bisimulation metric from~\cite{DesharnaisGJP04} (see also~\S{}\ref{subsubsec:introGameProbBisimMet}). This fact is not hard to check directly; one can also derive the coincidence via \Cref{example:KantorovichLiftingSubsumed} and the observations in~\cite{BaldanBKK18}.
\end{myexample}

\subsection{Joint Codensity Bisimulation}
\label{subsec:jointCodensityLifting}
We introduce a notion of \emph{joint codensity bisimulation}.
This minor variation of codensity bisimulation
becomes useful in the proof of soundness and completeness of our
game notion (\S{}\ref{sec:untrimmedGame}).

\begin{mydefinition}[joint codensity
  bisimulation]\label{def:jointCodensityBisim}
  Assume the setting of Def.~\ref{def:codensityBisimulation}.
  Let $\mathcal{V}\subseteq|\EE_X|$; joins in $\EE_X$ are denoted by
  $\bigsqcup$. We say that $\mathcal{V}$ is a {\em joint
    codensity bisimulation} over $c$ if
  $\bigsqcup_{P\in \mathcal{V}}P$ is a codensity bisimulation
  over $c$.
\end{mydefinition}
For instance, the set of all codensity bisimulations is a joint
codensity bisimulation, because the join $\nu\Phi_c^{\bOmega,\tau}$
is the largest bisimulation (a consequence of the Knaster--Tarski theorem).

\begin{mylemma}\label{lem:2050}
  In the setting of Def.~\ref{def:codensityBisimulation},  the
  downset $\mathop{\downarrow}(\nu\Phi_c^{\bOmega,\tau})$ is the largest
  joint codensity bisimulation (with respect to the inclusion order). \myqed
\end{mylemma}

\section{
  Untrimmed Games for Codensity Bisimilarity
}\label{sec:untrimmedGame}

As the first main technical contribution,  we introduce what we call the \emph{untrimmed} version of
codensity bisimilarity game. It is mathematically simple but its game
arenas can become much bigger than necessary. The \emph{trimmed}
version of games---with smaller arenas---will be introduced later in~\S{}\ref{sec:trimmedGames}, after
developing necessary categorical infrastructure.

Assume the setting of Def.~\ref{def:codensityBisimulation} for the rest of the section.

\begin{mydefinition}[untrimmed codensity bisimilarity game]\label{def:untrimmedCodensityGame}
  The
  \emph{untrimmed codensity bisimilarity game} is the safety game played
  by two players D and S, shown in
  Table~\ref{table:untrimmedCodensityGame}.
  \begin{table}[tbp]
    \caption{Untrimmed Codensity Bisimilarity Game}
    \label{table:untrimmedCodensityGame}
    \centering \renewcommand{\arraystretch}{1}
    \begin{tabular}{l|l|l}
      position & pl.\ &  possible moves    \\\hline\hline
      $P\in\EE_X$
               & S & $k\in\CC(X,\Omega)$ s.t.
      \\
               & & $\tau \circ F k \circ c:(X,P)\ndarrow(\Omega,\bOmega)$
      \\\hline
      $k\in\CC(X,\Omega)$
               & D &
                     $P'\in\EE_X$ s.t. $k:(X,P')\ndarrow(\Omega,\bOmega)$
    \end{tabular}
    \vspace{1.em}
    \caption{Untrimmed Codensity
      Game for Bisimulation
      Metric}
    \label{table:behMetricUntrimmedGame}
    \centering \renewcommand{\arraystretch}{1}
    \begin{tabular}{l|l|l}
      position & pl.\ &  possible moves    \\\hline\hline
      $d\in(\PMet)_X$
               & S & $k\in\Set(X,[0,1])$ s.t. \\
               & & $e \co F k \co c\not\in\PMet(d,d_{[0,1]})$
      \\\hline
      $k\in\Set(X,[0,1]])$
               & D & $d'\in(\PMet)_X$ s.t. \\
               & & $k\not\in\PMet(d',d_{[0,1]})$
    \end{tabular}
  \end{table}
\end{mydefinition}

\begin{mylemma}\label{lem:jointCodBisimIsInvSetUntr}
  Let
  $\mathcal{V}\subseteq|\EE_X|$. The following are
  equivalent.
  \begin{enumerate}
  \item $\mathcal{V}$ is an invariant for Duplicator
    (Def.~\ref{def:invariant}) in the safety game in
    Table~\ref{table:untrimmedCodensityGame}.
  \item $\mathcal{V}$ is a joint codensity bisimulation over
    $c$. \myqed
  \end{enumerate}
\end{mylemma}
\begin{mytheorem}
  The
  following coincide.
  \begin{enumerate}
  \item The set of all winning positions for D.
  \item The downset $\mathop{\downarrow}(\nu\Phi_c^{\bOmega,\tau})$ of
    the codensity bisimilarity.
  \end{enumerate}
\end{mytheorem}
\begin{myproof}
  On the one hand, the set of all winning positions for D is the
  largest invariant for D, by
  Prop.~\ref{prop:invariantsAndWinningPositions}.  On the other hand,
  the downset $\mathop{\downarrow}(\nu\Phi_c^{\bOmega,\tau})$ is the
  largest joint codensity bisimulation over $c$.  Thus, the statement
  follows from Lem.~\ref{lem:jointCodBisimIsInvSetUntr}.\myqed
\end{myproof}

We conclude that our game characterizes the codensity bisimilarity
$\nu \Phi_c^{\bOmega,\tau}$ (Def.~\ref{def:codensityBisimilarity}).
\begin{mycorollary}[soundness and completeness of untrimmed codensity
  games]
  \label{cor:soundnessAndCompletenessOfUntrimmedCodensityGame}
  $P\in \EE_{X}$ is a winning position for D if and only if
  $P\sqsubseteq \nu\Phi_c^{\bOmega,\tau}$.  \myqed
\end{mycorollary}

\begin{myexample}\label{example:behMetricUntrimmedGame}

  Recall Example \ref{example:behMetric}. Using the untrimmed
  codensity bisimilarity game, we can characterize the bisimulation
  metric from~\cite{DesharnaisGJP04}. Our general definition
  (Def.~\ref{def:untrimmedCodensityGame}) instantiates to the one in
  Table~\ref{table:behMetricUntrimmedGame}, which is however
  more complicated than the game we exhibited in the introduction
  (Table~\ref{table:probabilisticBisimMetricGameIntro}). For example, in
  Table~\ref{table:behMetricUntrimmedGame}, Duplicator's move is a 
  pseudometric $d\colon X^{2}\to[0,1]$ rather than a triple $(x,y,\varepsilon)$.

\end{myexample}

\section{Trimmed Codensity Games for Bisimilarity}
\label{sec:trimmedGames}

Our previous untrimmed game  (Table~\ref{table:untrimmedCodensityGame}) is pleasantly simple from a
theoretical point of view. However, as we saw in
Example~\ref{example:behMetricUntrimmedGame}, its instances tend to
have a much bigger arena than some known game notions.

Here we push our theory a step further, and present a fibrational
construction that allows us to \emph{trim} our games. We note that our construction still
remains on the fibrational level of abstraction.

\subsection{Generating Sets and Fibered Separators in a Fibration}
We start by building some fibrational infrastructure. The
following notion is a natural extension of the corresponding lattice-theoretic one.
Unlike in the case of algebraic lattices, we do not assume the
compactness of elements in $\mathcal{G}$.
\begin{mydefinition}[generating set]\label{def:generating set}
  Let $\fibp$ be a $\CLatw$-fibration and $X\in\CC$ be an object.  We say
  that a set $\mathcal{G}\subseteq|\EE_X|$ is a {\em generating set} of the
  fiber $\EE_X$ if, for any $P\in\EE_X$, there exists
  $\mathcal{A}\subseteq \mathcal{G}$ such that
  $\bigsqcup\mathcal{A}=P$.
\end{mydefinition}

\begin{myexample}\label{example:convBisimGen}
  Consider the $\CLatw$-fibration $\fib{\EqRel}{}{\Set}$ and $X\in\Set$. 
  For any $x,y\in X$, we define the equivalence relation $E_{x,y}$
  to be the least one equating $x,y$, that is,
  \begin{math}
    (z,w) \in E_{x,y}
  \end{math} if and only if
  \begin{math}
    (z=w \vee \{z,w\}=\{x,y\})
  \end{math}.
  Then the set
  \begin{math}
    \mathcal{G}=\{E_{x,y}~|~x,y\in X\}
  \end{math}
  of elements of the fiber $\EqRel_X$ is a generating set.
\end{myexample}
\begin{myexample}\label{example:behMetricGen}
  Recall Example \ref{example:behMetric}.  For
  $x,y\in X$ ($x\neq y$) and $r\in [0,1]$, 
  the  pseudometric $d_{x,y,r}$ over $X$ is defined by
  \begin{displaymath}\small
    d_{x,y,r}(z,w)=
    \begin{cases}
      0 & z=w \\
      r & \{z,w\}=\{x,y\} \\
      1 & \text{otherwise.}
    \end{cases}
  \end{displaymath}
  Then the set of pseudometrics $\{d_{x,y,r}~|~x,y\in X,x\neq y,r\in[0,1]\}$ is a
  generating set of the fiber $(\PMet)_X$.
\end{myexample}
One natural question is how to find such a generating set in the fiber over
the state space. Below we show that a generating set of the fiber of a special
object (called \emph{fibered separator}) induces a generating set of each
fiber by push-forward.
\begin{mydefinition}[fibered separator]\label{def:fibSep}
  Let $\fibp$ be a $\CLatw$-fibration. We say that $S\in\CC$ is a {\em
    fibered separator} if, for any $X\in\CC$ and $P,Q\in\EE_X$, we have
  \begin{displaymath}
    (\fa{f\in\CC(S,X)}f^*P=f^*Q)\implies P=Q.
  \end{displaymath}
\end{mydefinition}
\begin{mytheorem}\label{th:fibseptogen}
  Let $S\in\CC$ be a fibered separator of a $\CLatw$-fibration
  $\fibp$, and $\mathcal{G}\subseteq |\EE_S|$ be a generating set of
  $\EE_{S}$. For any $X\in\CC$, the following set is a generating set of
  $\EE_X$:
  \begin{displaymath}
    \{f_*P~|~P\in \mathcal{G},f\in\CC(S,X)\}.
  \end{displaymath}
  Here $f_{*}$ denotes the pushforward along $f$
  (\S\ref{subsec:fibrationPrelim}).
  \myqed
\end{mytheorem}

In fact, it was Thm.~\ref{th:fibseptogen} behind
Examples~\ref{example:convBisimGen}--\ref{example:behMetricGen}: in
both cases, $2\in\Set$ turns out to be a fibered separator for the
fibrations in question ($\fib{\EqRel}{}{\Set}$ and
$\fib{\PMet}{}{\Set}$), and the presented generating sets are obtained via
push-forward.

The following result  is useful in finding fibered
separators---see~\S{}\ref{subsec:quantitativeFKPMeas}.

\begin{myproposition}[change-of-base and fibered separators]
  \label{prop:baseChangeSep}
  Let $\fib{\EE}{p}{\CC}$ be a $\CLatw$-fibration, $R\colon\DD\to\CC$
  be a functor with a left adjoint $L\colon\CC\to\DD$, and $S\in\CC$ be a
  fibered separator for $p$. Then $LS\in\DD$ is a fibered separator of
  the change-of-base fibration $R^* p$. \myqed
\end{myproposition}

\subsection{$\mathcal{G}$-Joint Codensity Bisimulation}

We use generating sets to restrict moves in codensity  games.

\begin{mydefinition}
  In the setting of Def.~\ref{def:codensityBisimulation},  let
  $\mathcal{G}$ be a generating set of $\EE_X$.  A {\em $\mathcal{G}$-joint
    codensity bisimulation} over $c:X\to FX$ is a joint codensity
  bisimulation $\mathcal{V}$ over $c$ such that
  $\mathcal{V}\subseteq \mathcal{G}$.
\end{mydefinition}

\begin{mylemma}[key lemma]
  Assume the setting of Def.~\ref{def:codensityBisimulation}, and let
  $\mathcal{G}$ be a generating set of $\EE_X$.  The intersection
  $\bigl(\mathop{\downarrow}(\nu\Phi_c^{\bOmega,\tau})\bigr)\cap\mathcal{G}$ of
  the downset
  $\mathop{\downarrow}(\nu\Phi_c^{\bOmega,\tau})$ 
  and   the generating set $\mathcal{G}$
  is the largest
  $\mathcal{G}$-joint codensity bisimulation.
\end{mylemma}
\begin{myproof}
  Since $\mathcal{G}$ is a generating set, the union of all elements of
  $\mathop{\downarrow}(\nu\Phi_c^{\bOmega,\tau})\cap\mathcal{G}$ is
  equal to $\nu\Phi_c^{\bOmega,\tau}$.  Thus,
  $\mathop{\downarrow}(\nu\Phi_c^{\bOmega,\tau})\cap\mathcal{G}$ is a
  $\mathcal{G}$-joint codensity bisimulation.

  For any $\mathcal{G}$-joint codensity bisimulation $\mathcal{V}$, we
  have already shown
  $\mathcal{V}\subseteq\mathop{\downarrow}(\nu\Phi_c^{\bOmega,\tau})$.
  We also have $\mathcal{V}\subseteq\mathcal{G}$ by definition.  These imply
  $\mathcal{V}\subseteq\mathop{\downarrow}(\nu\Phi_c^{\bOmega,\tau})\cap\mathcal{G}$.\myqed
\end{myproof}

\subsection{Trimmed Codensity Bisimilarity Games}
The above structural results  lead  to our second game
notion.
\begin{mydefinition}[trimmed codensity bisimilarity
  game]\label{def:codensityBisimGame}
  Assume the setting of Def.~\ref{def:codensityBisimulation}, and that $\mathcal{G}\subseteq\EE_X$ is a generating set.  The
  \emph{codensity bisimilarity game} is the safety game played by two
  players D and S, shown in Table~\ref{table:codensityBisimGame}.

  \begin{table}[tbp]
    \caption{Trimmed Codensity Bisimilarity Game}
    \label{table:codensityBisimGame}
    \centering \renewcommand{\arraystretch}{1}
    \begin{tabular}{l|l|l}
      position & pl.\ &  possible moves    \\\hline\hline
      $P\in \mathcal{G}$
               & S & $k\in\CC(X,\Omega)$ s.t. \\
               & & $\tau \circ Fk \circ c:(X,P)\ndarrow(\Omega,\bOmega)$
      \\\hline
      $k\in\CC(X,\Omega)$
               & D & $P'\in \mathcal{G}$ s.t. \\
               & & $k:(X,P')\ndarrow(\Omega,\bOmega)$
    \end{tabular}
  \end{table}

\end{mydefinition}

Assume the setting of Def.~\ref{def:codensityBisimGame} for the rest of the section.

\begin{mylemma}\label{lem:jointCodBisimIsInvSet}
  Let $\mathcal{V}\subseteq|\EE_X|$.  The
  following are equivalent:
  \begin{enumerate}
  \item $\mathcal{V}$ is an invariant for D (Def.~\ref{def:invariant}) in the game in
    Table~\ref{table:codensityBisimGame}.
  \item $\mathcal{V}$ is a $\mathcal{G}$-joint codensity bisimulation
    over $c$. \myqed
  \end{enumerate}
\end{mylemma}

\begin{mytheorem}
  The following sets coincide.
  \begin{enumerate}
  \item The set of D-winning positions  in the game in
    Table~\ref{table:codensityBisimGame}.
  \item The intersection
    $\bigl(\mathop{\downarrow}(\nu\Phi_c^{\bOmega,\tau})\bigr)\cap \mathcal{G}$ of
    the downset     of the
    codensity bisimilarity over $c$ and the generating set
    $\mathcal{G}$. \myqed
  \end{enumerate}
\end{mytheorem}

We conclude that our second game characterizes the codensity
bisimilarity $\nu \Phi_c^{\bOmega,\tau}$
(Def.~\ref{def:codensityBisimilarity}) too.
\begin{mycorollary}[soundness and completeness of trimmed codensity
  games]
  \label{cor:soundnessAndCompletenessOfCodensityGame}
  In Def.~\ref{def:codensityBisimGame},
  $P\in \mathcal{G}$ is a winning position for Duplicator if and only
  if $P\sqsubseteq \nu\Phi_c^{\bOmega,\tau}$.
  \myqed
\end{mycorollary}

\section{Multiple Observation Domains}
\label{sec:multipleObservationDomains}

We extend the theory so far and accommodate multiple observation domains and modalities. This extension is needed for some examples, such as those marked with $\dagger$ in Table~\ref{table:codensityLiftings}. 

We consider the class $\Lift(F,p)$ of liftings of an endofunctor
$F:\CC\arrow\CC$ along a $\CLatw$-fibration $\fibp$. It comes with a
natural pointwise partial order:
\begin{equation}\label{eq:orderBetweenLifting}
  G\sqsubseteq H\iff\fa{X\in\EE}GX\sqsubseteq HX\quad
  (G,H\in\Lift(F,p)),
\end{equation}
and the partially ordered class $\Lift(F,p)$ admits meets of arbitrary size.
As done in the original codensity lifting of endofunctors in
\cite{SprungerKDH18} (and monads in \cite{KatsumataS15}), we extend
the codensity lifting so that it takes a family of parameters
$\{(\bOmega_A,\tau_A)\}_{A\in\AAA}$, and returns the {\em
  intersection} of the codensity liftings of $F$ with these
parameters. 
\begin{mydefinition}[codensity lifting of a
  functor with multiple parameters~\cite{SprungerKDH18}]\label{def:multCodensityLifting}

  Let $F:\CC\arrow\CC$ be a functor,
  $\fibp$ be a $\CLatw$-fibration,
  $\AAA$ be a class, and
  $\{(\bOmega_A,\tau_A)\}_{A\in\AAA}$ be an $\AAA$-indexed family of parameters (of the codensity lifting of $F$ along $p$),
  which is denoted simply  by $(\bOmega,\tau)$.
  The (multiple-parameter) \emph{codensity lifting} of $F$ with $(\bOmega,\tau)$ is the endofunctor
  $F^{\bOmega,\tau}:\EE\arrow\EE$ defined by the intersection of the
  codensity liftings:
  \begin{align*}
    F^{\bOmega,\tau} P
    & = \bigsqcap_{A\in\AAA}F^{\bOmega_A,\tau_A}P, \\
    & \text{that is, } \bigsqcap_{A\in\AAA, k \in \EE (P, \bOmega_A)}
      \bigl(\tau_A \circ F (p(k))\bigr)^{\ast} (\bOmega_A).
  \end{align*}
\end{mydefinition}

The rest of the theoretical development is completely parallel to the one in the
previous sections. 
Therefore we only present key definitions and the
main result (Cor.~\ref{cor:multSoundnessAndCompletenessOfCodensityGame}). The omitted definitions and results can be recovered from the ones in \S{}\ref{sec:codensityBisimilarity}--\ref{sec:trimmedGames}, by replacing a single-parameter codensity lifting (Def.~\ref{def:codensityLifting}) by a multi-parameter one (Def.~\ref{def:multCodensityLifting}).

\begin{mydefinition}[codensity bisimulation]
  \label{def:multCodensityBisimulation}
  Assume the setting of Def.~\ref{def:multCodensityLifting}.  Let
  $c:X\to FX$ be an $F$-coalgebra.  An object $P\in\EE_X$ is a {\em
    codensity bisimulation} over $c$ if
  $c:(X,P) \darrow (FX,F^{\bOmega,\tau} P)$; that is, $c\colon X\to FX$
  is decent with respect to the designated indistinguishability
  structures.
\end{mydefinition}

\begin{mydefinition}[codensity bisimilarity game]\label{def:multCodensityBisimGame}
  In the setting of Def.~\ref{def:multCodensityBisimulation}, let
  $\mathcal{G}$ be a generating set of $\EE_X$.
  The \emph{codensity bisimilarity game} is the safety game, played by
  two players D and S, shown in Table
  \ref{table:multCodensityBisimGame}.

  \begin{table}[tbp]
    \caption{Trimmed Codensity Bisimilarity Game with Multiple
      Observations}
    \label{table:multCodensityBisimGame}
    \centering \renewcommand{\arraystretch}{1}
    \begin{tabular}{l|l|l}
      position & pl.\ &  possible moves    \\\hline\hline
      $P\in \mathcal{G}$
               & S & $A\in\AAA$ and $k\in\CC(X,\Omega_A)$ s.t. \\
               & & $\tau_A \circ F k \circ c:(X,P)\ndarrow(\Omega_A,\bOmega_A)$
      \\\hline
      $A\in\AAA$ and
               & D & $P'\in \mathcal{G}$ s.t.\\
      $k\in\CC(X,\Omega_A)$
               & & $k:(X,P')\ndarrow(\Omega_A,\bOmega_A)$
    \end{tabular}
  \end{table}

\end{mydefinition}

\begin{mycorollary}[soundness and completeness of codensity games]
  \label{cor:multSoundnessAndCompletenessOfCodensityGame}
  Assume the setting of Def.~\ref{def:multCodensityBisimGame}.
  $P\in \EE_{X}$ is a winning position for Duplicator if and only if
  $P\sqsubseteq \nu\Phi_c^{\bOmega,\tau}$.   \myqed
\end{mycorollary}

\begin{myexample}[bisimulation topology for deterministic automata]\label{example:bisimTop}
  Here we describe the topological example in~\Cref{table:fibex}. 
  Consider the $\CLatw$-fibration $\fib{\Top}{}{\Set}$ and the
  functor $A_\Sigma=2\times(\place)^\Sigma\colon\Set\to\Set$, where $\Sigma$ is a fixed alphabet. Coalgebras for this functor are  deterministic automata over $\Sigma$; see e.g.~\cite{Jacobs16coalgBook,Rutten00a}. 

  We take the following data as a parameter of codensity lifting (cf.\ Def.~\ref{def:multCodensityLifting}): 
  $\AAA=\{\varepsilon\}\cup\Sigma$, 
  $\bOmega_\alpha$ is the Sierpinski space for each $\alpha\in \AAA$, and the modalities
  $\tau_\varepsilon, \tau_a\colon A_\Sigma 2\to 2$ (where  $a\in \Sigma$) are defined by
  \begin{displaymath}
    \tau_\varepsilon(t,\rho)=t\quad\text{and}\quad
    \tau_a(t,\rho)=\rho(a).
  \end{displaymath}
  Recall that the Sierpinski space is the set $2=\{\bot,\top\}$  with the topology $\{\emptyset,\{\top\},2\}$; this observation domain models the situation where acceptance of a word is only
  \emph{semi}-decidable.

  Let $c\colon X\to A_\Sigma X$ be a deterministic automata.
  The above choice of parameters leads to the following  codensity bisimilarity:
  the state space $X$ is equipped with the topology generated by the following family of open sets.
  \begin{displaymath}
    \{x\in X\mid \text{$w$ is accepted from  $x$} \} \subseteq X,\quad
    \text{for each } w\in \Sigma^{*} 
  \end{displaymath}
  One can extract various information from this \emph{bisimulation topology} via standard topological constructs. For example, the specialization order of this topology coincides with the language inclusion order.

  For illustration by comparison, consider changing the observation domain
  from the Sierpinski space to the discrete 2-point set.
  The bisimulation topology over $X$ is now generated by
  \begin{displaymath}
    \begin{array}{l}
      \{x\in X\mid \text{$w$ is accepted from $x$} \} 
      \text{ and }
      \\
      \{x\in X\mid \text{$w$ is not accepted from $x$} \},\quad
      \text{for each }w\in \Sigma^{*}. 
    \end{array}
  \end{displaymath}
  We can now  observe rejection of a word, too, because $\{\bot\}\subseteq 2$ is open. The specialization order of this topology is the language equivalence, and it satisfies the R0 separation axiom (while the last Sierpinski example does not). 

  We take these examples of bisimulation topology as
  a process-semantical incarnation of the ``observability via topology, computability via continuity'' paradigm from domain theory. 
  The definition of codensity bisimulation (cf.\ Def.~\ref{def:codensityLifting}) fits well with this intuition, too: a continuous map $k\colon (X,P)\darrow\bOmega$ in Def.~\ref{def:codensityLifting} is a ``computable observation''; accordingly, an open set of the bisimulation topology is a property that is decided by finitely many of  those computable observations.

\end{myexample}

\section{Transfer of Codensity Bisimilarities}
\label{sec:relatingDifferentSit}

In our formulation, for the same endofunctor $F\colon\CC\to\CC$, we can use
various $\CLatw$-fibrations and parameters $(\bOmega, \tau)$ to equip $F$-coalgebras with different
bisimilarity-like notions.  Some relations among those  codensity
bisimilarities can be categorically captured by the following
theorem.

\begin{mytheorem}[transfer of codensity bisimilarity]
  \label{prop:transferOfBisim}
  Let $\fib{\EE}{p}{\CC}$ and $\fib{\FF}{q}{\CC}$ be
  $\CLatw$-fibrations, $F\colon\CC\to\CC$ be an endofunctor,
  $c\colon X\to FX$ be an $F$-coalgebra,
  $T\colon\EE\to\FF$ be a full and faithful fibered functor from $p$ to $q$ preserving
  fibered meets, and $\{(\bOmega_A,\tau_A)\}_{A\in\AAA}$ be an
  $\AAA$-indexed family of parameters for codensity lifting of $F$
  along $p$.  
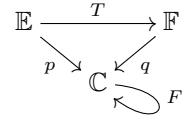
  \begin{displaymath}
    \vcenter{\xymatrix@R=1em@C=1.5em{
        {\EE}
        \ar[rr]^-{T}
        \ar[rd]_{p}
        &&
        {\FF}
        \ar[ld]^{q}
        \\
        &
        {\CC}
        \ar@(r,rd)[]^(.4){F}
      }} 
  \end{displaymath}
  In this setting, $\{(T\bOmega_A,\tau_A)\}_{A\in\AAA}$ is an
  $\AAA$-indexed family of parameters for codensity lifting of $F$
  along $q$, and we have
  $\nu\Phi_c^{T\bOmega,\tau}=T(\nu\Phi_c^{\bOmega,\tau})$. \myqed
\end{mytheorem}

\begin{myexample}
  \label{example:convBisimTransfer}
  We show that the codensity bisimilarities in
  Examples~\ref{ex:eqrelpowlift} \&~\ref{ex:erelpowlift} are indeed
  the usual bisimilarity notions for Kripke frames.  Recall that they
  are build on the two $\CLatw$-fibrations $\fib{\EqRel}{}{\Set}$ and
  $\fib{\ERel}{}{\Set}$.

  We first note that the inclusion functor $i\colon\EqRel\to\ERel$ is
  a reflection, having the equivalence closure
  $(\place)^{\mathrm{eq}}\colon\ERel\to\EqRel$ as the left adjoint. It
  follows that $i$ is meet-preserving.  Moreover, $i$ is fibered.

  \begin{displaymath}
    \vcenter
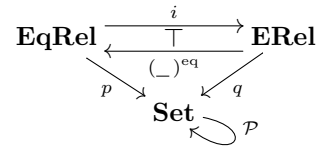
{\xymatrix@R=1.5em@C=1.5em{
        {\EqRel}
        \ar[rd]_{p}
        &&
        {\ERel}
        \adjunction{ll}{(\place)^{\mathrm{eq}}}{i}
        \ar[ld]^{q}
        \\
        &
        {\Set}
        \ar@(r,rd)[]^(.4){\pow}
      }} 
  \end{displaymath}
  We introduce shorthands $\dot\pow_2,\dot\pow_3$ for the liftings in
  Examples~\ref{ex:eqrelpowlift} \&~\ref{ex:erelpowlift}:
  \begin{displaymath}
    \begin{array}{rclll}
      \dot\pow_2&=&\pow^{\Eq_2,\diamond}&:\EqRel\arrow\EqRel&\text{(Example~\ref{ex:eqrelpowlift})},\\
      \dot\pow_3&=&\pow^{\Eq_2,\diamond}&:\ERel\arrow\ERel&\text{(Example~\ref{ex:erelpowlift})}
    \end{array}
  \end{displaymath}
  Now, for the sake of our proof, let us introduce a relational lifting
  $\dot\pow_1:\ERel\arrow\ERel$ of $\pow$ along $\fib{\ERel}{}{\Set}$,
  for which it is obvious that the corresponding bisimilarity notion is
  the usual bisimilarity for Kripke frames. We do so in concrete terms,
  instead of as a codensity lifting:
  \begin{align*}\small
    \begin{array}{rcl}
      (S,T)\in\dot\pow_1(R) & \Longleftrightarrow &
                                                    (\fa{x\in S}\ex{y\in T}(x,y)\in R ) ~ \wedge \\
                            & & (\fa{y\in T}\ex{x\in S}(x,y)\in R).
    \end{array}  
  \end{align*}
  We note that $\dot\pow_2$ is the restriction of $\dot\pow_1$ from
  $\ERel$ to $\EqRel$ along $i$. Note also that
  $\dot\pow_3= \dot\pow_1\circ i\circ (\place)^{\mathrm{eq}}$.

  Let $c:X\arrow\pow X$ be a Kripke frame and
  $\Phi_i=c^*\circ \dot\pow_i$ ($i=1,2,3$) be the predicate transformer
  corresponding to each lifting.  Theorem \ref{prop:transferOfBisim}
  states that $\nu\Phi_3=i(\nu\Phi_2)$.

  Furthermore, by $\dot\pow_1\sqsubseteq\dot\pow_3$ (where $\sqsubseteq$
  is the order in~(\ref{eq:orderBetweenLifting})), we have
  $\nu\Phi_1\sqsubseteq\nu\Phi_3$.  From
  $i\circ\dot\pow_2=\dot\pow_1\circ i$ and fiberedness of $c$, we can see
  that $i(\nu\Phi_2)$ is a fixed point of $\Phi_1$, which yields
  $i(\nu\Phi_2)\sqsubseteq\nu\Phi_1$ by the Knaster--Tarski theorem. The
  three (in)equalities so far allow us to
  conclude $\nu\Phi_3=i(\nu\Phi_2)=\nu\Phi_1$, stating that the
  conventional bisimilarity $\nu\Phi_1$ is equal to the codensity
  bisimilarities in Examples~\ref{ex:eqrelpowlift}
  \&~\ref{ex:erelpowlift}. As a consequence, the conventional
  bisimilarity $\nu\Phi_1$ is necessarily an equivalence relation.
\end{myexample}

\section{Examples}\label{sec:examples}

\subsection{ Kripke Frames and (Conventional) Bisimilarity}\label{subsec:exampleConventionalBisimilarity}

We consider $\fib{\EqRel}{}{\Set}$ as an underlying
$\CLatw$-fibration, and a Kripke frame $c:X\arrow\pow X$
(as a $\pow$-coalgebra). We further use
the codensity lifting
$\pow^{\Eq_2,\diamond}$ (Example
\ref{ex:eqrelpowlift}) and
the generating set described
in Example \ref{example:convBisimGen} to trim games; the resulting game is
shown in Table~\ref{table:exampleConventionalBisimilarity}. As shown
in Example \ref{example:convBisimTransfer}, the codensity bisimilarity
$\nu\Phi_c^{\Eq_2,\diamond}$ (which is $\nu\Phi_2$ in \Cref{example:convBisimTransfer}) coincides with conventional bisimilarity on
$c$ using the standard relational lifting of $\pow$ to $\ERel$.

\begin{table}[htbp]
  \caption{Codensity Bisimilarity Game for Conventional Bisimilarity}
  \label{table:exampleConventionalBisimilarity}
  \centering \renewcommand{\arraystretch}{1}
  \begin{tabular}{l|l|l}
    position & pl.\ &  possible moves    \\\hline\hline
    $(x,y)\in X\times X$
             & S &
                   $k\in\Set(X,2)$ s.t.\
    \\
             & &
                 \quad$\ex{x'\in c(x)}k(x')=\top$ \\
             & &
                 \quad$\not\Leftrightarrow \ex{y'\in c(y)}k(y')=\top$ \\\hline
    $k\in\Set(X,2)$
             & D &
                   $(x'',y'')$ s.t.\
                   $k(x'') \neq k(y'')$
  \end{tabular}

  \vspace{1em}
  \caption{Codensity Bisimilarity Game for Deterministic Automata and
    Their Language Equivalence}
  \label{table:DFA}
  \centering \renewcommand{\arraystretch}{1}
  \begin{tabular}{l|l|l}
    position & pl.\ &  possible moves    \\\hline\hline
    $(x,y)\in X\times X$
             & S &
                   If $\pi_1(x)\neq\pi_1(y)$ then S wins
    \\
             & &
		 If $\pi_1(x)=\pi_1(y)$ then
    \\
             & & \quad $a\in\Sigma$ and $k\in\Set(X,2)$ \\
             & & \quad s.t.\ $k(\pi_2(x)(a))\neq k(\pi_2(y)(a))$ \\\hline
    $a\in\Sigma$ and
             & D &
                   $(x'',y'')\in X\times X$ s.t.\
                   $k(x'')\neq k(y'')$ \\
    $k\in\Set(X,2)$ & &
  \end{tabular}

  \vspace{1em}

  \caption{Codensity Bisimilarity Game for Deterministic Automata and
    Bisimulation Topology}
  \label{table:DFATopology}
  \centering \renewcommand{\arraystretch}{1}
  \begin{tabular}{l|l|l}
    position & pl.\ &  possible moves    \\\hline\hline
    $\mathcal{O}\in \Top_X$
             & S &
                   $a\in\{\varepsilon\}\cup\Sigma$ and $k\in\Set(X,2)$
    \\
             & &
		 s.t.\ $\tau_a \circ (A_\Sigma k) \circ c:(X,\mathcal{O})\ndarrow(2,\bOmega_a)$\\\hline
    $a\in\{\varepsilon\}\cup\Sigma$
             & D &
                   $\mathcal{O}'\in \Top_X$ \\
    and $k\in\Set(X,2)$ & & s.t.\ $k:(X,\mathcal{O}')\ndarrow(2,\bOmega_a)$
  \end{tabular}

  \vspace{1em}

  \caption{Codensity Bisimilarity Game for Nondeterministic Automata
    and Their Bisimilarity}
  \label{table:NFA}
  \centering \renewcommand{\arraystretch}{1}
  \begin{tabular}{l|l|l}
    position & pl.\ &  possible moves    \\\hline\hline
    $(x,y)\in X\times X$
             & S &
                   If $\pi_1(x)\neq\pi_1(y)$ then S wins
    \\
             & &
		 If $\pi_1(x)=\pi_1(y)$ then
    \\
             & & \quad$a\in\Sigma$ and $k\in\Set(X,2)$ \\
             & & \quad s.t.\ $\ex{x'\in\pi_2(x)(a)}k(x')=\top$ \\
             & & \quad\quad$\nLeftrightarrow\ex{y'\in\pi_2(y)(a)}k(y')=\top$ \\\hline
    $a\in\Sigma$ and
             & D &
                   $(x'',y'')\in X\times X$ s.t.\
                   $k(x'')\neq k(y'')$ \\
    $k\in\Set(X,2)$ & &
  \end{tabular}

  \vspace{1em}

  \caption{Codensity Bisimilarity Game for Probabilistic Bisimilarity}
  \label{table:FKP}
  \centering \renewcommand{\arraystretch}{1}
  \begin{tabular}{l|l|l}
    position & pl.\ &  possible moves    \\\hline\hline
    $(x,y)$
             & S &
                   $r\in[0,1]$ and $k\in\Set(X,2)$ s.t.\
    \\
    $\in X\times X$ & & $c(x)(k^{-1}(\top))\ge r > c(y)(k^{-1}(\top))$, or \\
             & & $c(y)(k^{-1}(\top))\ge r > c(x)(k^{-1}(\top))$ \\\hline
    $r\in[0,1]$ and
             & D &
                   $(x'',y'')$ s.t.\ $k(x'') \neq k(y'')$ \\
    $k\in\Set(X,2)$ & &
  \end{tabular}

\end{table}

\begin{mytheorem}
  $(x,y)$ is a D-winning position if and only if
  $(x,y)\in\nu\Phi_c^{\Eq_2,\diamond}$, if and only if $x$ and $y$ are
  bisimilar. \myqed
\end{mytheorem}

Extension of this result to labeled transition systems and Kripke models (with valuations)
is straightforward, using suitable choice of endofunctors---see
also~\S{}\ref{subsec:DFA}.

\subsection{Deterministic Automata and Their Language
  Equivalence}\label{subsec:DFA}
We use the functor $A_{\Sigma}\colon\Set\to\Set$ from Example~\ref{example:bisimTop}, for which a coalgebra is a deterministic automaton.
Let us lift
$A_{\Sigma}$ along the fibration $\fib{\EqRel}{}{\Set}$, with the same modalities
$\tau_\varepsilon, \tau_a\colon A_\Sigma 2\to 2$ as in Example~\ref{example:bisimTop} (where  $a\in \Sigma$). Our observation domains are $\bOmega_{\alpha}=(2,\Eq_{2})$  for all $\alpha\in \{\varepsilon\}\cup\Sigma$. 
The resulting codensity lifting
$(A_\Sigma)^{\bOmega,\tau}\colon\EqRel\to\EqRel$ is concretely described
as follows.
\begin{align*}\footnotesize
  \begin{array}{ll}
    &(A_\Sigma)^{\bOmega,\tau}(R)= \\
    &\left\{
      \begin{array}{l|l}
        \big((t_1,\rho_1), & \fa{k\colon X\to2} \\
        (t_2,\rho_2)\big) & (\fa{x,y\in X}(x,y)\in R\Rightarrow k(x)=k(y)) \\
        \in(A_\Sigma X)^2 &\Rightarrow (t_1=t_2) \wedge \\
                           & (\fa{a\in\Sigma} (k\co \rho_1)(a) = (k\co \rho_2)(a) )
      \end{array}
                             \right\}.
  \end{array}
\end{align*}
Let $c\colon X\to A_\Sigma X$ be a
deterministic automaton. 
It is not hard to see that the codensity bisimilarity
$\nu\Phi_c^{\bOmega,\tau}$ coincides with
language equivalence of deterministic automata.  Our trimmed codensity game is shown in Table~\ref{table:DFA} (in a slightly optimized form). The game therefore characterizes the language
equivalence (Cor.~\ref{cor:multSoundnessAndCompletenessOfCodensityGame}).

\subsection{Deterministic Automata and  the Language Topology}\label{subsec:DFAtopology}
We introduced two versions of \emph{bisimulation topology} for deterministic automata in Example~\ref{example:bisimTop}. They are in close correspondences with accepted languages; therefore we call them \emph{language topologies}. 

For the first topology in Example~\ref{example:bisimTop} (where $\bOmega$ is the Sierpinski space, capturing that acceptance is only semi-decidable), the corresponding (untrimmed) codensity game is shown in Table~\ref{table:DFATopology}. It follows from our general results that the game notion is sound and complete.

We have not yet found a good way (e.g.\ generating sets) of trimming the
game arena; this is left as future work.

\subsection{Nondeterministic Automata and  Bisimilarity}\label{subsec:NFA}
Let us now turn to nondeterministic automata, that is,
$N_\Sigma$-coalgebras for the functor $N_\Sigma=2\times (\pow \place)^{\Sigma}$.
Much like the situation for DFAs, we lift this functor along the
$\CLatw$-fibration $\fib{\EqRel}{}{\Set}$ by codensity lifting with multiple
observation domains, as follows.  Let $\AAA$ be the set
$\{\varepsilon\}\cup\Sigma$.  We set the parameter of codensity lifting
as follows, where $a\in\Sigma$.
\begin{displaymath}
  \bOmega_\varepsilon=\bOmega_a=\Eq_2,\;
  \tau_\varepsilon(t,\rho)=t,\;
  \tau_a(t,\rho)=\diamond(\rho(a)).
\end{displaymath}
The resulting codensity lifting
$(N_\Sigma)^{\bOmega,\tau}\colon\EqRel\to\EqRel$ is
concretely described as
\begin{align*}\footnotesize
  \begin{array}{ll}
    &(N_\Sigma)^{\bOmega,\tau}(R)= \\
    &\left\{
      \begin{array}{l|l}
        \big((t_1,\rho_1), & \fa{k\colon X\to2} \\
        (t_2,\rho_2)\big) & (\fa{x,y\in X}(x,y)\in R\Rightarrow k(x)=k(y)) \\
        \in(N_\Sigma X)^2 &\Rightarrow (t_1=t_2) \wedge \\
                           & \qquad\big(\fa{a\in\Sigma} \top \in (k\co \rho_1)(a) \\
                           & \qquad\quad \Leftrightarrow \top \in (k\co \rho_2)(a) \big)
      \end{array}
                             \right\}.
  \end{array}
\end{align*}
Let $c\colon X\to N_\Sigma X$ be a nondeterministic automaton.  
It is again not hard to see that the codensity bisimilarity
$\nu\Phi_c^{\bOmega,\tau}$ is the usual notion of bisimilarity of nondeterministic automata.  Our trimmed codensity game
is shown in Table~\ref{table:NFA}, in a slightly optimized form, and it captures bisimilarity.

A topological variant of the above story is possible, much like
in~\S{}\ref{subsec:DFAtopology}.

\subsection{Markov Chains and Bisimulation Metric}
\label{subsec:quantitativeFKP}

Recall Examples \ref{example:behMetric},
\ref{example:behMetricUntrimmedGame}, and
\ref{example:behMetricGen}. Markov chains are $\sdist$-coalgebras.
We use the $\CLatw$-fibration $\fib{\PMet}{}{\Set}$, taking
pseudometrics as a notion of indistinguishability.  With the lifting parameter
we described in Example \ref{example:behMetric}, we get the
bisimulation metric as the codensity bisimilarity.  We can use the
generating set described in Example \ref{example:behMetricGen} to obtain a
trimmed codensity game; the resulting game essentially coincides with
the one in Table~\ref{table:probabilisticBisimMetricGameIntro} in the
introduction. Therefore,
Cor.~\ref{cor:soundnessAndCompletenessOfCodensityGame} gives an
abstract proof for the correctness of the game.

\subsection{Continuous State Markov Chains and Bisimulation Metric}
\label{subsec:quantitativeFKPMeas}
In order to accommodate continuous state Markov chains (for
which measurable structures are essential), we consider an example
that involves $\Meas$.  Continuing~\S{}\ref{subsec:quantitativeFKP},
by the change-of-base along the forgetful functor
$U\colon\Meas\to\Set$, we get another $\CLatw$-fibration
$\fib{U^*(\PMet)}{}{\Meas}$. A continuous state Markov chain is a
coalgebra $X\to \mathcal{G}_{\le 1}X$ of the so-called \emph{sub-Giry}
functor over $\Meas$---see, e.g.,~\cite{Hasuo15CMCSJournVer}.

Since the forgetful functor $\Meas\to\Set$ has a left adjoint,
Prop.~\ref{prop:baseChangeSep} gives us a fibered separator for
$U^*(\PMet)\to\Meas$.  This gives us a game notion similar to that
in~\S{}\ref{subsec:quantitativeFKP}.

\subsection{Markov Chains and Probabilistic Bisimilarity }
\label{subsec:FKPGame}
In order to define bisimilarity-like equivalence relation on Markov
chains, we first lift $\sdist$ along the $\CLatw$-fibration
$\fib{\EqRel}{}{\Set}$. For that purpose, here we use the following
multiple lifting parameters.  The index set is $\AAA=[0,1]$.  For each
$r\in\AAA$, we set $\bOmega_r=(2,\Eq_2)$, and define a \emph{threshold modality}
$\tau_r\colon \sdist 2\to2$ by
$\tau_r(p)=\top$ if and only if $p(\top)\ge r$. Then for any $R\in\EqRel_X$, the
relation part of the codensity lifting $\sdist^{\bOmega,\tau}(X,R)$
relates $p,q\in\sdist(X)$ if and only if
\begin{align*}
  & \small\fa{r\in[0,1]} \fa{k\colon X\to2}
    \bigl((\fa{(x,y)\in R} k(x) = k(y))\bigr. \\
  & \small\textstyle
    \qquad\bigl.\implies\bigl(\sum_{x\in k^{-1}(\top)}p(x)\ge r\Leftrightarrow\sum_{x\in k^{-1}(\top)}q(x)\ge r\bigr)\bigr).
\end{align*}

Let us fix a Markov chain $c\colon X\to \sdist X$.  All
these data give rise to $\sdist^{\bOmega,\tau}$ and
$\nu\Phi_c^{\bOmega,\tau}$ as in Definitions
\ref{def:multCodensityBisimulation} and
\ref{def:codensityBisimilarity}. It is not hard to see that the resulting codensity bisimilarity coincides with probabilistic bisimilarity in~\cite{LarsenS91}. Note, for example, that a relation-preserving map $k\colon (X,R)\darrow (2,\Eq_{2})$ coincides with an $R$-closed subset of $X$. 
The resulting trimmed codensity game is  in
Table~\ref{table:FKP}. It is essentially the same as
Table~\ref{table:probabilisticBisimGameIntroFKP} (arising
from~\cite{FijalkowKP17}).  The difference is that $r$ is additionally
present in Table~\ref{table:FKP}; it is easy to realize that $r$ plays
no role in the game.

\section{Conclusions and Future Work}
Motivated by some recent
works~\cite{FijalkowKP17,KoenigM18,BonchiKP18,BaldanBKK18}, and
especially by the similarity of the two games
(Tables~\ref{table:probabilisticBisimGameIntroFKP}
and~\ref{table:probabilisticBisimMetricGameIntro}), we introduced a
fibrational framework that uniformly describes the correspondence
between various bisimilarity notions and games. The fibrational
abstraction allows us to accommodate some new examples, such as
bisimulation topology. Moreover, the structural theory developed
in~\S{}\ref{sec:multipleObservationDomains}--\ref{sec:relatingDifferentSit}
provides new insights to the nature of bisimilarity, we believe, identifying the crucial role of observation maps ($k\colon X\to \Omega$ in Def.~\ref{def:codensityLifting}) in bisimulation notions. 

As future work, we intend to accommodate \emph{modal logics} as is done
in~\cite{KoenigM18}. We are also interested in using games with more
complex winning conditions (e.g.\ parity); they have been used for
(bi)simulation notions for B\"{u}chi and parity
automata~\cite{EtessamiWS05}. Finally, we will pursue the algorithmic
use of the current results.

\IEEEtriggeratref{18}
\bibliographystyle{IEEEtran}\bibliography{IEEEabrv,bibliography}

\newpage
\appendix

\subsection{Direct Proof of Equivalence of the Two Game Notions Characterizing Probabilistic Bisimilarity (Tables~\ref{table:probabilisticBisimGameIntroFKP},~\ref{table:probabilisticBisimGameDesharnais})}
\label{appendix:gameEquivDirectProofProbBisim}

\subsubsection{Table~\ref{table:probabilisticBisimGameDesharnais} $\leadsto$
  Table~\ref{table:probabilisticBisimGameIntroFKP} }
Assume that Duplicator wins Table~\ref{table:probabilisticBisimGameDesharnais}  from $(x,y)$, and let Spoiler play some $Z$ in Table~\ref{table:probabilisticBisimGameIntroFKP} . There are two cases to consider which are essentially identical, but we write them down separately just to make sure.
\begin{itemize}
\item If $\tau(x,Z)>\tau(y,Z)$ then make Spoiler select $s=x$ and play $Z$ in Table~\ref{table:probabilisticBisimGameDesharnais}. To this Duplicator responds with some $Z'\supseteq Z$ such that $\tau(x,Z)\leq \tau(y,Z')$, which implies that $Z'\neq Z$. Pick any $y'\in Z'\setminus Z$ and play it as Spoiler in Table~\ref{table:probabilisticBisimGameDesharnais}; when Duplicator responds with some $x'\in Z$, play the pair $x'$ and $y'$ as Duplicator in Table~\ref{table:probabilisticBisimGameIntroFKP}.
\item If $\tau(x,Z)<\tau(y,Z)$ then make Spoiler select $s=y$ and play $Z$ in Table~\ref{table:probabilisticBisimGameDesharnais}. To this Duplicator responds with some $Z'\supseteq Z$ such that $\tau(y,Z)\leq \tau(x,Z')$, which implies that $Z'\neq Z$. Pick any $y'\in Z'\setminus Z$ and play it as Spoiler in Table~\ref{table:probabilisticBisimGameDesharnais}; when Duplicator responds with some $x'\in Z$, play the pair $x'$ and $y'$ as Duplicator in Table~\ref{table:probabilisticBisimGameIntroFKP}.
\end{itemize}

\subsubsection{
  Table~\ref{table:probabilisticBisimGameIntroFKP}
  $\leadsto$
  Table~\ref{table:probabilisticBisimGameDesharnais}
}

This is a less straightforward implication. A winning strategy for Duplicator in Table~\ref{table:probabilisticBisimGameDesharnais}  is built not from a single strategy in Table~\ref{table:probabilisticBisimGameIntroFKP}, but rather from an entire collection of winning positions.

Formally, assume that Duplicator wins Table~\ref{table:probabilisticBisimGameIntroFKP}  from $(x,y)$, and let Spoiler choose $s\in\{x,y\}$ and play some $Z$ in Table~\ref{table:probabilisticBisimGameDesharnais}. Define
\[
  \bar{Z} = \{w\in X \mid \exists v\in Z \mbox{ s.t.~Duplicator wins Table~\ref{table:probabilisticBisimGameIntroFKP}  from }(v,w)\}.
\]
One basic observation is that $Z\subseteq\bar{Z}$, since Duplicator wins from all positions of the form $(w,w)$. As a result:
\begin{equation}\label{eq:xZleqxbZ}
  \tau(x,Z)\leq \tau(x,\bar{Z}) \qquad \mbox{and} \qquad \tau(y,Z)\leq \tau(y,\bar{Z}).
\end{equation}
Another observation is that Spoiler wins Table~\ref{table:probabilisticBisimGameIntroFKP} from the position $\bar{Z}$. To see this, consider any Duplicator's response $x'\in\bar{Z}$, $y'\not\in\bar{Z}$. Then there is some $v\in Z$ such that Duplicator wins Table~\ref{table:probabilisticBisimGameIntroFKP} from $(v,x')$. If Duplicator could win Table~\ref{table:probabilisticBisimGameIntroFKP}  from $(x',y')$ then she could win from $(v,y')$ as well, which contradicts the assumption that $y'\not\in\bar{Z}$.

Since we assume that Duplicator wins Table~\ref{table:probabilisticBisimGameIntroFKP}  from $(x,y)$, $\bar{Z}$ cannot be a legal move for Spoiler from $(x,y)$, hence
\[
  \tau(x,\bar{Z})=\tau(y,\bar{Z}).
\]
Together with~\eqref{eq:xZleqxbZ} this implies that
\[
  \tau(x,Z)\leq \tau(y,\bar{Z}) \qquad \mbox{and} \qquad \tau(y,Z)\leq \tau(x,\bar{Z}),
\]
so $Z'=\bar{Z}$ is a legal move for Duplicator in stage (ii) of Table~\ref{table:probabilisticBisimGameDesharnais}, no matter if Spoiler chose $s=x$ or $s=y$ in stage (i). To this, in stage (iii) replies with some $y'\in \bar{Z}\setminus Z$. By definition of $\bar{Z}$, there is some $v\in Z$ such that Duplicator wins Table~\ref{table:probabilisticBisimGameIntroFKP} from $(v,y')$, so Duplicator can respond with $x'=v$.

\subsection{Introduction to $\CLatw$-Fibration}
\label{appendix:CLatwFib}
We present an introduction to ($\CLatw$-)fibrations, starting from a functor $F_{\EE}\colon\CC^{\op}\to \CLatw$. The relevance of the latter is explained in~\S{}\ref{subsec:fibrationPrelim}. For details, readers are referred to~\cite{Jacobs99a}.

\subsubsection{The Grothendieck Construction}
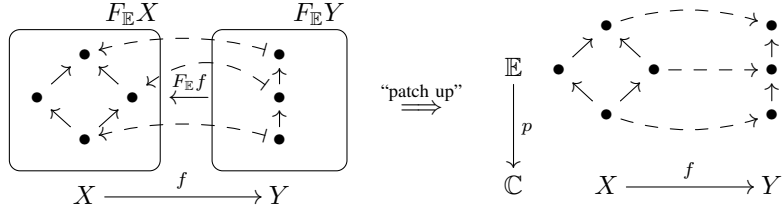
\begin{figure*}[tbp]
  \begin{align*}
    \def\fiber#1{\save
    [].[ddrr]!C="fiber#1"*+<1.5em>[F-:<4pt>]\frm{}\restore}    \def\fibername#1{\save[]+<0em,.5em>*{#1}\restore}    \vcenter{\xymatrix@R=.2em@C=.5em{
          &&\fibername{F_{\EE} X} &&&&\fibername{F_{\EE} Y}\\
    \fiber1           & {\bullet} &           && \fiber2      & {\bullet} \ar@/_/@{|-->}[llll] &   \\
    {\bullet}\ar[ru] &           & {\bullet}\ar[lu] &
                                                      \stackrel{F_{\EE} f}{\longleftarrow}
          &                                   & {\bullet}\ar[u] \ar@/_1.3em/@{|-->}[lll] &
    \\
          & {\bullet}\ar[lu]\ar[ru] &           &    & & {\bullet}\ar[u]
                                                         \ar@/_/@{|-->}[llll]
                                    &
    \\
    \\
          & X \ar[rrrr]^{f}
                           &&&& Y
                                }}
                                \quad\stackrel{\text{``patch up''}}{\Longrightarrow}\quad
                                \vcenter{\xymatrix@R=.5em@C=.5em{
          & & {\bullet} &           &&
                                  & {\bullet} \ar@/^/@{-->}[llll];[] &   \\
    {\EE}\ar[ddd]^{p}&
                       {\bullet}\ar[ru] &           & {\bullet}\ar[lu] &
                                  & {\phantom{\bullet}} & {\bullet}\ar[u] \ar@/^0em/@{-->}[lll];[] &
                                                                                                     {\phantom{\bullet}}
    \\
          && {\bullet}\ar[lu]\ar[ru] &           &    & & {\bullet}\ar[u]
                                                          \ar@/_/@{-->}[llll];[]
                                    &
    \\
    \\
    {\CC} && X \ar[rrrr]^{f}
                           &&&& Y
                                }}
  \end{align*}
  \caption{The Grothendieck construction}
  \label{fig:GrothendieckConstrSketch}
\end{figure*}

In general, the equivalence between index categories $\CC^{\op}\to \mathbf{Cat}$ and
fibrations is well-known. Here we sketch the \emph{Grothendieck
  construction} from the former to the latter, focusing the special case of $\CC^{\op}\to \CLatw$ and $\CLatw$-fibrations. Its idea is to ``patch up''
the family  $\bigl(F_{\EE}X\bigr)_{X\in\CC}$ of complete lattices, and form a big category $\EE$, as shown in Fig.~\ref{fig:GrothendieckConstrSketch}.

On the right-hand side in Fig.~\ref{fig:GrothendieckConstrSketch},  we add some arrows (denoted by $\dashrightarrow$)  so
that we have an arrow
$(F_{\EE} f)(Q)\to Q$ in $\EE$ for each $Q\in F_{\EE} Y$.
(On the left-hand side, the correspondence
$\raisebox{.3pt}{$\shortmid$}\!\joinrel\dashrightarrow$ depicts the
action of the map $F_{\EE} f$.) The  diagram in $\EE$ in Fig.~\ref{fig:GrothendieckConstrSketch}
should be understood as a Hasse diagram:
those arrows which arise from composition are not depicted.

\begin{mydefinition}[The Grothendieck construction]\label{definition:GrothendieckConstr}
  Given $F_{\EE}\colon \CC^{\op}\to\CLatw$, we define the category $\EE $
  by
  \begin{itemize}
  \item its objects: a pair $(X,P)$ of an object $X\in\CC$ and an
    element $P$ of the poset $F_{\EE} X$;  and
  \item its arrows: $f\colon (X,P)\to (Y,Q)$ is an arrow $f\colon X\to Y$ in $\CC$ such
    that
    \begin{displaymath}
      P\sqsubseteq (F_{\EE} f)(Q).
    \end{displaymath}
    Here
    $\sqsubseteq$ refers
    to the order of $F_{\EE} X$.
  \end{itemize}
\end{mydefinition}

Thus arises a category $\EE$ that incorporates:
\begin{itemize}
\item the order structure of each of the posets $(F_{\EE} X)_{X\in\CC}$, and
\item the pullback
structure by $(F_{\EE} f)_{f\colon\text{$\CC$-arrow}}$. 
\end{itemize}
For fixed $X\in\CC$, the
objects of the form $(X,P)$ and the arrows $\id_{X}$ between them form a
subcategory of $\EE$. This is denoted by $\EE_{X}$ and called the
\emph{fiber} over $X$. It is obvious that $\EE_{X}$ is a poset
that is isomorphic to $F_{\EE} X$.

Moreover, there is a canonical projection functor $p\colon \EE\to\CC$ that
carries $(X,P)$ to $X$.

\subsubsection{Formal Definition of $\CLatw$-Fibration}
We axiomatize those structures which arise in the way described above.
\begin{mydefinition}[$\CLatw$-fibration]  A \emph{$\CLatw$-fibration} $\fibp$ consists of two categories $\EE,\CC$
  and a functor $p\colon \EE\to\CC$, that satisfy the following properties.
  \begin{itemize}
  \item Each fiber $\EE_{X}$ is a complete lattice. Here the \emph{fiber} $\EE_{X}$ for
    $X\in\CC$ is the subcategory of $\EE$ consisting of the following data: objects
    $P\in\EE$
    such that $pP=X$; and arrows $f\colon P\to Q$ such that
    $pf=\id_{X}$ (such arrows are said to be \emph{vertical}).
  \item
    Given $f\colon X\to Y$ in $\CC$ and $Q\in \EE_{Y}$, there is an object
    $f^{*} Q\in \EE_{X}$ and an $\EE$-arrow $\overline{f}Q\colon f^{*}Q\to Q$ with
    the following universal property. For any $P\in \EE_{X}$ and $g\colon P\to Q$
    in $\EE$, if $pg=f$ then $g$ factors through $\overline{f}(Q)$
    uniquely via a vertical arrow. That is, there exists
    unique $g'$ such that $g=\overline{f}(Q)\co g'$ and $pg'=\id_{X}$.
    \begin{displaymath}
      \vcenter{\xymatrix@R=1.2em{
          {\EE}
          \ar[dd]^{p}
          &
          &
          {Q}
          \ar@{}[rrdd]|{\Longrightarrow}
          &&
          {f^{*}Q}
          \ar[r]^-{\overline{f}(Q)}
          &
          {Q}
          \\
          &&&&
          {P}
          \ar[ru]_-{g}
          \ar@{-->}[u]^{g'}
          \\
          {\CC}
          &
          {X}
          \ar[r]^{f}
          &
          {Y}
          &&
          {X}
          \ar[r]^{f}
          &
          {Y}
        }}
    \end{displaymath}
  \item The correspondences $(\place)^{*}$ and $\overline{(\place)}$ are
    functorial:
    \begin{align*}
      \id_{Y}^{*}Q&=Q\enspace,&
                                (g\co f)^{*}(Q) &= f^{*}(g^{*}Q),\\
      \overline{\id_{Y}}(Q)&=\id_{Q}\enspace,&
                                               \overline{g\co f}(Q) &= \overline{g}Q\co \overline{f}(g^{*}Q).
    \end{align*}
    The last equality can be depicted as follows.
    \begin{displaymath}
      \vcenter{\xymatrix@R=.4em{
          {\EE}
          \ar[ddd]^{p}
          &
          {f^{*}(g^{*}Q)}
          \ar[r]^-{\overline{f}(g^{*}Q)}
          \ar@{=}[d]
          &
          {g^{*}Q}
          \ar[r]^{\overline{g}Q}
          &
          {Q}
          \\
          &
          {(g\co f)^{*}Q}
          \ar@/_/[rru]_{\overline{g\co f}(Q)}
          &&
          \\\\
          {\CC}
          &
          {X}
          \ar[r]^-{f}
          &
          {Y}
          \ar[r]^{g}
          &
          {Z}
        }}
    \end{displaymath}
  \end{itemize}
  The category $\EE$ is called the \emph{total category} of the fibration;
  $\CC$ is the \emph{base category}. The arrow
  $\overline{f}Q\colon f^{*}Q\to Q$ is called the \emph{Cartesian lifting} of
  $f$ and $Q$.   An arrow in $\EE$ is \emph{Cartesian} (or \emph{reindexing}) if it coincides with
  $\overline{f}Q$ for some $f$ and $Q$.
\end{mydefinition}
In the case where $\fibp$ is induced by an indexed category
$F_{\EE}\colon \CC^{\op}\to\CLatw$ via Def.~\ref{definition:GrothendieckConstr},
a
Cartesian lifting is given by $f^{*}(Q)=(F_{\EE} f)(Q)$.

In the current paper we focus on $\CLatw$-fibrations. In a (general) fibration, a fiber $\EE_{X}$ is
not just a preorder but a category, and this elicits a lot of technical
subtleties.
Nevertheless, it should  not be hard to generalize the current paper's observations
to general, not necessarily $\CLatw$-, fibrations (especially to the split ones).
We shall often denote a vertical arrow in $\EE$ (i.e.\ an arrow inside a fiber)
by $\sqsubseteq$.

\subsection{Codensity Characterization of Hausdorff pseudometric}
\label{subsec:hausdorff}

\begin{myproposition}
  Let $(X,d)$ be a pseudometric space.
  For any $S,T\subseteq X$, we define two functions \[
    d_H(S,T)=\max\left(\sup_{x\in S}\inf_{y\in T} d(x,y),\sup_{y\in T}\inf_{x\in S} d(x,y)\right)
  \] and \[
    d_c(S,T)=\sup_{k\in\PMet((X,d),([0,1],d_{\mathbb{R}}))}d_{\mathbb{R}}\left(\inf_{x\in S}k(x),\inf_{y\in T}k(y)\right).
  \]

  The values of two functions coincide.
\end{myproposition}
\begin{myproof}
  First, we show $d_c(S,T)\ge d_H(S,T)$ by contradiction.

  Suppose it doesn't hold.
  Then, by definition, at least one of \[
    \sup_{x\in S}\inf_{y\in T} d(x,y)
  \] and \[
    \sup_{y\in T}\inf_{x\in S} d(x,y)
  \] is greater than $d_c(S,T)$.
  We can assume the former is greater than $d_c(S,T)$ w.l.o.g.

  Therefore, for some $x_0 \in S$, \[
    d_c(S,T) < \inf_{y\in T} d(x_0,y)
  \] holds.

  Now, since $d(x_0,\place)$ is a non-expansive function by the triangle inequality,
  we have \[
    d_c(S,T) \ge d_{\mathbb{R}}\left(\inf_{x\in S}d(x_0,x),\inf_{y\in T}d(x_0,y)\right).
  \]
  However, since $\inf_{x\in S}d(x_0,x)=0$, we have $d_c(S,T) \ge \inf_{y\in T}d(x_0,y)$, which is a contradiction.

  Next, we show $d_c(S,T) \le d_H(S,T)$ by contradiction.

  Suppose $d_c(S,T) > d_H(S,T) + \varepsilon$ for some $\varepsilon > 0$.
  Then, for some non-expansive $k\colon X\to[0,1]$, \[
    d_{\mathbb{R}}\left(\inf_{x\in S}k(x),\inf_{y\in T}k(y)\right) > d_H(S,T) + \varepsilon
  \] holds.

  W.l.o.g. we can assume $\inf_{x\in S}k(x) \le \inf_{y\in T}k(y)$.

  Thus, for some $x_0 \in S$ and $y_0 \in T$ satisfying $k(x_0) \le \inf_{x\in S}k(x) + \varepsilon/5$ and $k(y_0) \le \inf_{y\in T}k(y) + \varepsilon/5$, \[
    d_{\mathbb{R}}(k(x_0),k(y_0)) > d_H(S,T) + 3\varepsilon/5
  \] holds.
  Since \[
    d_H(S,T) \ge \sup_{x\in S}\inf_{y\in T} d(x,y),
  \]
  there exists some $y_1 \in T$ satisfying \[
    d_H(S,T) \ge d(x_0,y_1) \ge d_{\mathbb{R}}(k(x_0),k(y_1)).
  \]

  However, we have $k(x_0) \le k(y_0) + \varepsilon/5 \le k(y_1) + 2\varepsilon/5$, so \[
    d_{\mathbb{R}}(k(x_0),k(y_1) + \varepsilon/5) \ge d_{\mathbb{R}}(k(x_0),k(y_0) + 2\varepsilon/5)
  \] and \[
    d_{\mathbb{R}}(k(x_0),k(y_1)) + 3\varepsilon/5 \ge d_{\mathbb{R}}(k(x_0),k(y_0))
  \] holds.

  Then,
  \begin{align*}
    & d_{\mathbb{R}}(k(x_0),k(y_0)) \\
    & \le d_{\mathbb{R}}(k(x_0),k(y_1)) + 3\varepsilon/5 \\
    & \le d_H(S,T) + 3\varepsilon/5 \\
    & < d_{\mathbb{R}}(k(x_0),k(y_0))
  \end{align*}
  holds, which is a contradiction.\myqed
\end{myproof}

\subsection{Omitted Proofs}
\subsubsection{Proof of Lem.~\ref{lem:2050}}

\begin{myproof}
  The downset $\mathop{\downarrow}(\nu\Phi^{\bOmega,\tau})$ is a joint codensity bisimulation, because the union of all elements of $\mathop{\downarrow}(\nu\Phi^{\bOmega,\tau})$ is equal to a codensity bisimulation $\nu\Phi^{\bOmega,\tau}$.

  Let $\mathcal{V}$ be a joint codensity bisimulation. Then for any $P\in \mathcal{V}$,
  we have $P\sqsubseteq\nu\Phi^{\bOmega,\tau}$, because $P\sqsubseteq\bigsqcup_{Q\in \mathcal{V}}Q\sqsubseteq\nu\Phi^{\bOmega,\tau}$.\myqed
\end{myproof}

\subsubsection{Proof of Lem.~\ref{lem:jointCodBisimIsInvSetUntr}}
\begin{myproof}
  We follow the following logical equivalence.
  \begin{align*}
    1)\iff&\left(
            \parbox{.7\linewidth}{\small
            $\fa{P\in \mathcal{V},k\in \CC(X,\Omega)}
            \tau \co F k \co c \not\in \EE(P,\bOmega)
            $ \\ $\implies \ex{P'\in \mathcal{V}}
    k\not\in \EE(P',\bOmega)$
    }
    \right)\\
    \iff&\left(
          \parbox{.7\linewidth}{\small
          $\fa{P\in \mathcal{V},k\in \CC(X,\Omega)}$\\
    $
    (\fa{P'\in \mathcal{V}}k\in \EE(P',\bOmega))$
    \\$
    \implies
    \tau \co F k \co c\in \EE(P,\bOmega)$
    }
    \right)\\
    \iff&\left(
          \parbox{.7\linewidth}{\small
          $\fa{k\in \CC(X,\Omega)}
          (k\in \EE(\bigsqcup_{P'\in \mathcal{V}}P',\bOmega))
          \implies
          \tau \co F k \co c\in \EE(\bigsqcup_{P\in \mathcal{V}}P,\bOmega)$
          }
          \right)\\
    \iff
          &\textstyle
            \bigsqcup_{P\in \mathcal{V}}P\sqsubseteq\Phi^{\bOmega,\tau}(\bigsqcup_{P'\in \mathcal{V}}P'),
  \end{align*}
  where the last equivalence is by Thm.~\ref{thm:codensityBisimChara}.
\end{myproof}

\subsubsection{Proof of Prop.~\ref{prop:baseChangeSep}}
\begin{myproof}
  For any $X\in\DD$, the bijection $\DD(LS,X)\cong\CC(S,RX)$ exists.
  Thus, naturally identifying $(R^*\EE)_X$ and $\EE_{RX}$, we have the following  for any $P,Q\in(R^*\EE)_X$.
  \begin{align*}\small
    \begin{array}{ll}
      &\fa{f\in\DD(LS,X)} f^* P = f^* Q \\
      &\implies \fa{f\in\DD(LS,X)} (Rf)^* P = (Rf)^* Q \\
      &\implies \fa{f\in\DD(LS,X)} (Rf\co \eta_S)^* P = (Rf\co \eta_S)^* Q \\
      &\implies \fa{g\in\CC(S,RX)} g^* P = g^* Q \\
      &\implies P=Q    \end{array}
  \end{align*}

  \vspace{-1.3em}
  \myqed
\end{myproof}

\subsubsection{Proof of Prop.~\ref{prop:transferOfBisim}}
\begin{myproof}

  We have $T\Phi^{\bOmega,\tau}=\Phi^{T\bOmega,\tau}T$
  because, for any $P\in\FF_X$,
  \begin{align*}
    &T\Phi^{\bOmega,\tau}P \\
    &=T\left( \bigsqcap_{A\in\AAA} \bigsqcap_{k \in \FF (P, \bOmega(A))}
      (\tau_A \circ F (r k) \circ c)^{\ast} \bOmega(A) \right)\\
    &=\bigsqcap_{A\in\AAA} \bigsqcap_{k \in \FF (P, \bOmega(A))}
      (\tau_A \circ F (r k) \circ c)^{\ast} T\bOmega(A) \\
    &=\bigsqcap_{A\in\AAA} \bigsqcap_{k \in \FF (P, \bOmega(A))}
      (\tau_A \circ F (r k) \circ c)^{\ast} T\bOmega(A) \\
    &=\bigsqcap_{A\in\AAA} \bigsqcap_{k \in \FF (P, \bOmega(A))}
      (\tau_A \circ F (p (T k)) \circ c)^{\ast} T\bOmega(A) \\
    &=\bigsqcap_{A\in\AAA} \bigsqcap_{l \in \EE (TP, T\bOmega(A))}
      (\tau_A \circ F (p l) \circ c)^{\ast} T\bOmega(A) \\
    &=\Phi^{T\bOmega,\tau}TP
  \end{align*}
  holds.

  Considering this and the fact that $T$ preserves meets,
  \cite[Lemma 20]{SprungerKDH18} implies
  $T(\nu\Phi^{\bOmega,\tau})=\nu\Phi^{T\bOmega,\tau}$.\myqed
\end{myproof}

\end{document}